\documentclass[reprint,floatfix,showpacs]{revtex4-2}

\usepackage{graphicx}
\usepackage{dcolumn}
\usepackage{amsmath}
\usepackage{amssymb}
\usepackage{color}

\begin{document}

\title{The Chaotic Dynamics of Coupled Maps with Convection and Diffusion}
\author{A. Raj}
\author{M. R. Paul}
\email{Corresponding author: mrp@vt.edu}
\affiliation{Department of Mechanical Engineering, Virginia Tech, Blacksburg, Virginia 24061}

\date{\today}

\begin{abstract}

We explore the fundamental role of convective and diffusive coupling on the spatiotemporal chaos of a large one-dimensional lattice of coupled maps on a ring.  Extensively chaotic dynamics persist over nearly the entire range of spatial coupling strengths that yield numerically stable dynamics. The spatial coupling affects the shape of the spectrum of Lyapunov exponents, for increasing convective coupling the leading Lyapunov exponents increase while the large negative Lyapunov exponents decrease resulting in a significant increase in the fractal dimension of the dynamics. Convective coupling increases the localization of the leading Lyapunov vector and adds significant power to the high-wavenumber components of the spectrum of covariant Lyapunov vectors while increasing their entanglement. The eigenvalues and eigenvectors of the diffusive coupling matrix remain insightful quantities in the presence of nonlinear convective coupling. We discuss the implications of these findings for high-dimensional chaotic dynamics of large transport systems.
\end{abstract}

\maketitle

\section{Introduction}
\label{section:introduction}
Many important problems of interest are spatially extended high-dimensional systems that are driven far-from-equilibrium to yield dynamics that are spatiotemporally chaotic~\cite{cross:1993}. In many of these systems, transport phenomena connect the dynamics between spatial regions. Convective and diffusive transport mechanisms are ubiquitous in nature. Examples include the complex motion of the atmosphere~\cite{pedlosky:1987,stephan:2025}, atmospheric rivers~\cite{gimeno:2014}, the formation and dynamics of clouds~\cite{velarde:1980,chavez-medina:2026}, oceanic circulation~\cite{marshall:1999,wunsch:2004}, convective patterns in the sun's photosphere~\cite{ahlers:2009:solar}, the dynamics of moving populations~\cite{holmes:1994} and reaction-advection-diffusion systems~\cite{dewit:2020,mukherjee:2022} to name a few.

One important source of complexity is the nonlinear convective term that appears in many transport equations.  For example, the convective term in the Navier-Stokes equation and its fundamental connection with fluid turbulence~\cite{pope:2000,xiao:2019}. Building a physical understanding of the influence of convection and diffusion on high-dimensional chaotic dynamics remains a central challenge in the study of nonequilibrium systems. In many of these systems, the essential physical idea is that spatiotemporal chaos is the result of locally generated disorder that is distributed spatially by linear and nonlinear transport mechanisms~\cite{hohenberg:1989}.

Fundamental questions are often explored using model systems that are more numerically accessible.  A widely used model to study spatiotemporal dynamics is the coupled map lattice (CML)~\cite{bunimovich:1988,kaneko:1992,kaneko:1989-stc}. In a CML, local disorder generated by a nonlinear mapping function is transported throughout the lattice by spatial coupling terms that can be tailored as desired. In our study, we focus on the role of convection and diffusion with the transport equations of momentum, energy, and chemical species in mind. 

Advances in dynamical systems theory continue to yield deep insights into the fundamental nature of chaos. Analyzing a system's trajectory through its state-space, and the growth or decay of small perturbations in its tangent-space, have greatly improved our understanding of chaotic dynamics. Advances that can be generalized to high-dimensional dynamics include unstable periodic orbit theory ~\cite{budanur:2015,cleary:2025}, Koopman mode decomposition~\cite{mezic:2013,williams:2015}, dynamic mode decomposition~\cite{tu:2014,schmid:2022}, exact coherent structures~\cite{waleffe:2001,kawahara:2012}, and covariant Lyapunov vectors (CLVs)~\cite{pikovsky:2016,ginelli:2007}.  

Quantifying the dynamics of the CLVs has emerged as a powerful approach yielding new insights into chaos~\cite{kuptsov:2012,pikovsky:2016}. Using CLVs, the degree of hyperbolicity of the dynamics can be estimated, and for several important model systems, the tangent-space can be decomposed into physical and transient modes~\cite{takeuchi:2011,yang:2013}. The CLVs have been used to estimate the angle between stable and unstable manifolds of Kolmogorov flow~\cite{inubushi:2012}, the hyperbolicity properties of Rayleigh-B\'enard convection~\cite{xu:2016,xu:2018} and Hamiltonian systems~\cite{pradoreynoso:2021}, and to explore localization properties in recurrent neural networks~\cite{engelken:2023}.

We previously investigated the role of diffusive coupling~\cite{raj:2024}, and an extended spatial linear coupling~\cite{raj:2026}, on the chaotic dynamics of CMLs. When the spatial coupling is linear, the lattice dynamics can be represented in terms of a coupling matrix. The eigenvalues of the coupling matrix can be used to predict the spectrum of Lyapunov exponents and, consequently, the fractal dimension~\cite{takeuchi:2011,raj:2024,raj:2026}. However, when nonlinear convective coupling is present, this eigen-representation is no longer possible and a theoretical description is not currently available. Exploring the role and influence of nonlinear spatial coupling is central to our investigation.

The remainder of the paper is organized as follows. In \S\ref{section:approach}, we describe our general approach and the convective and diffusive spatial coupling we use with the CMLs. Our results are described in \S\ref{section:results}. We analyze the spacetime dynamics with drift in \S\ref{section:drift}, the variation of the spectrum of Lyapunov exponents with the convective and diffusive coupling strengths in \S\ref{section:lyapunov-exponents}, and the dimension of the attractor describing the dynamics in \S\ref{section:fractal-dimension}. The relationship between the CLVs and the eigenvectors of the diffusive coupling matrix is explored in \S\ref{section:clvs} and the role of spatial coupling on the entanglement of the CLVs is quantified in \S\ref{section:vdos}. Our conclusions are presented in \S\ref{section:conclusion}.

\section{Approach}
\label{section:approach}

\subsection{Coupled Map Lattices with Convection and Diffusion}
\label{section:cmls}

We use a spatially periodic 1D lattice of coupled-maps with convective and diffusive coupling with dynamics that are described by 
\begin{multline}
u_i^{(n+1)} = f(u_i^{(n)}) + \frac{\epsilon}{2} \left[ f(u_{i+1}^{(n)}) - 2 f(u_i^{(n)}) + f(u_{i-1}^{(n)}) \right] \\ - \frac{\alpha}{2} f(u_i^{(n)}) \left[f(u_{i+1}^{(n)})-f(u_{i-1}^{(n)}) \right]
\label{eq:cml}
\end{multline}
where $u_i$ is the scalar quantity of interest, $n$ is the integer time step, and $i$ is the lattice index that ranges over $i\!=\! 1, 2, \ldots, N$  where $N$ is the total number of lattice sites. The strength of the diffusive and convective couplings are given by $\epsilon$ and $\alpha$, respectively.

The convective and diffusive couplings are spatially symmetric and involve the nearest neighbors about each lattice site. The spatial coupling terms, represented by the second and third terms on the right hand side of Eq.~\eqref{eq:cml}, can be thought of as the result of a central difference operator applied to a diffusive and convective term, respectively, in a transport equation. We emphasize that it is not our intention to model the physics of a particular differential equation, rather our intention is to study the role of diffusive and convective type coupling terms on chaotic dynamics.  We consider lattices with periodic boundary conditions such that $u_i^{(n)} = u_{i+N}^{(n)}$ for all $i$ and $n$.

The mapping function $f(u)$  in Eq.~\eqref{eq:cml} generates local disorder. In our study, we focus upon the quadratic map 
\begin{equation}
f(u) = r \left( \frac{1}{4} - u^2 \right),
\label{eq:quadratic-map}
\end{equation}
where $r \!>\! 0$ is the control parameter. It will be useful to first briefly describe the dynamics of an isolated map before discussing the numerical stability limits of the lattice dynamics. 

For $r\!>\!0$, the quadratic map has the two fixed points $u_{\pm} = (2 r)^{-1} \left( - 1 \pm \sqrt{r^2 + 1} \right)$ where $u_+$ is stable for $0 \!<\! r \!\le\! \sqrt{3}$, unstable for $r \!>\! \sqrt{3}$, and $u_-$ is unstable for $r>0$. With increasing $r$, the quadratic map undergoes a period doubling route to chaos~\cite{may:1976,ott:1993} with chaos first appearing at $r_\infty \approx 2.37$. For chaotic dynamics, $r \!\ge\! r_\infty$, the values of $u$ are between $u_\text{max} \!=\! f(0) \!=\! r/4$ and $u_\text{min} \!=\! f(r/4)$. The lower limit is found by iterating the map one step from its maximum value.

For chaotic dynamics, the fixed point $u_+$ lies within the region containing the chaotic attractor, $u_\text{min} \!\le\! u_+ \le u_\text{max}$. At $r \!=\! r_\infty$, the fixed point $u_-$ lies outside of the chaotic attractor, $u_- \!<\! u_\text{min}$, and it repels dynamics that approach it from above back into the chaotic regime. However, as $r$ increases, $u_-$ approaches $u_{\text{min}}$ and eventually collides with the attractor. This results in a boundary crisis~\cite{grebogi:1982} annihilating the chaotic attractor when $u_- \!=\! f(r/4)$ which occurs at $r_c \!=\! 2 \sqrt{2}$. For $r>r_c$, the values of $u$ diverge to negative infinity.  As a result, the chaotic dynamics of a single map are limited to $r_\infty \!\le\! r \!<\! r_c$.

\begin{figure}
\begin{center}
\includegraphics[width=1in]{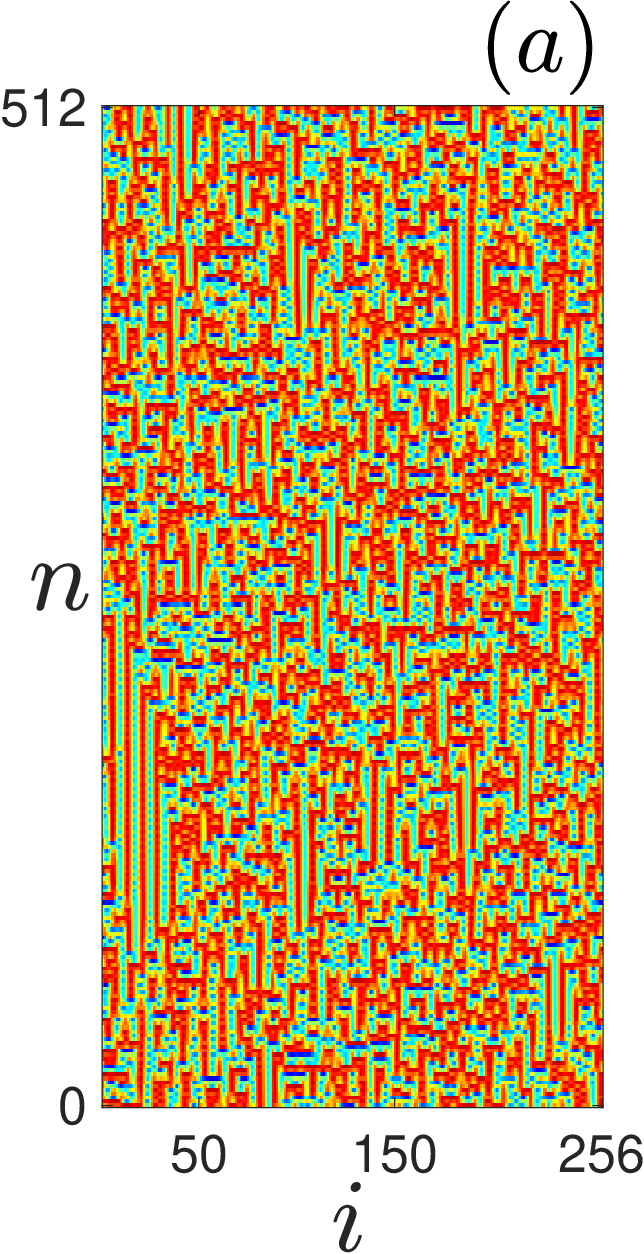} \hspace{0.1cm}
\includegraphics[width=1in]{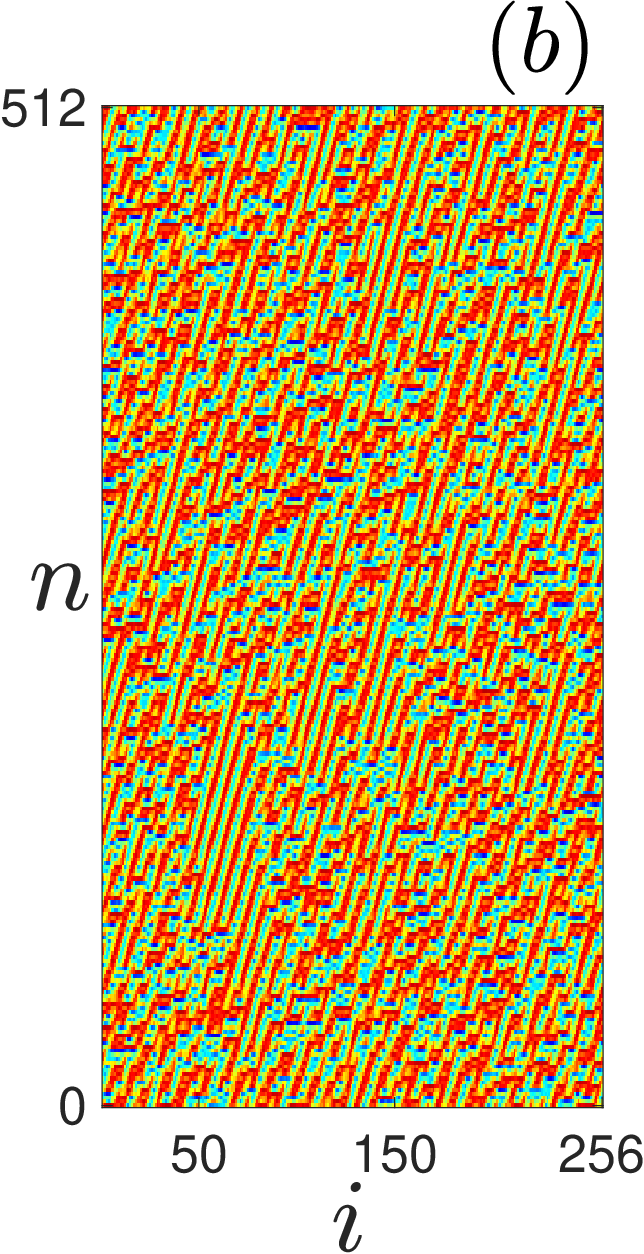} \hspace{0.1cm}
\includegraphics[width=1in]{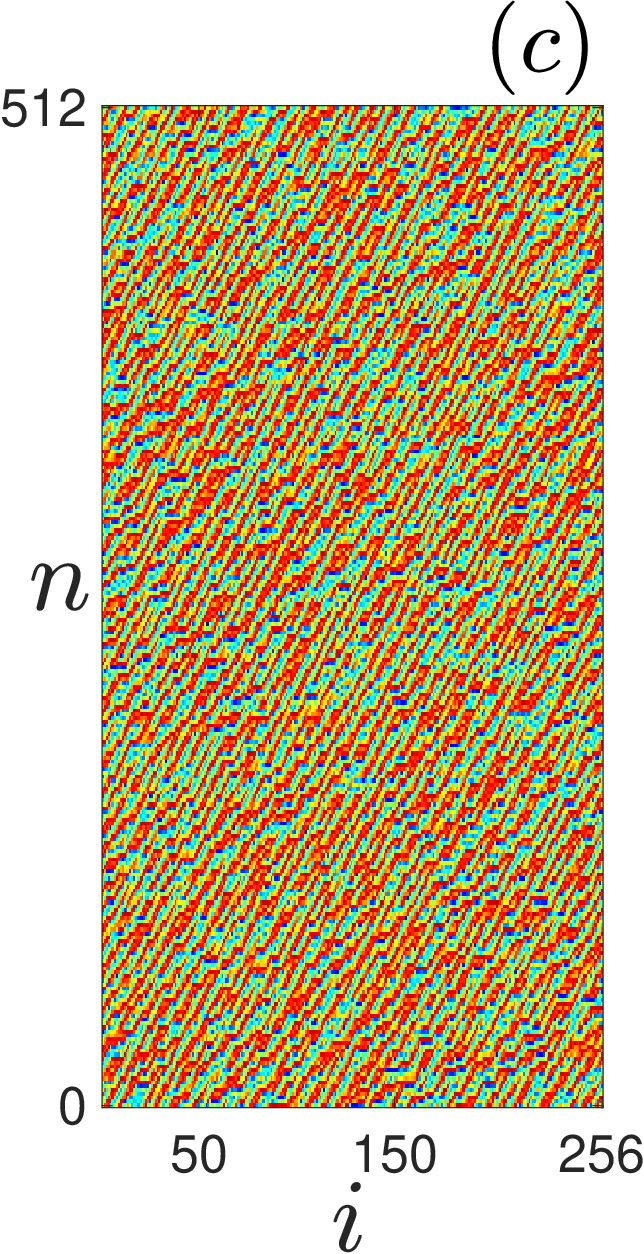} \hspace{0.1cm} \\ \vspace{0.2cm}
\includegraphics[width=2.75in]{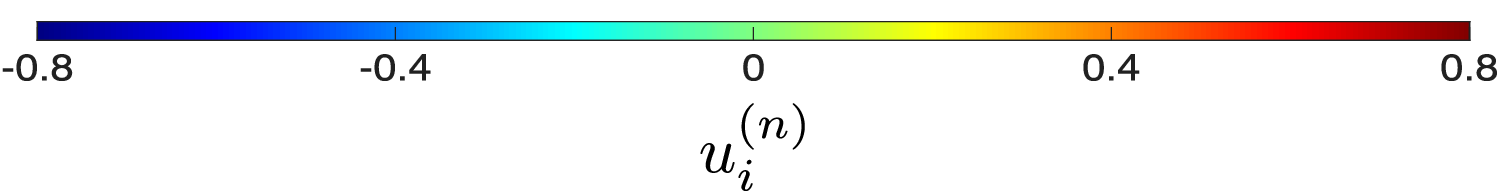}
\end{center}
\vspace{-9pt}
\caption{Spacetime dynamics of $u_i^{(n)}$ shown for alternating time-steps: ($a$)~$\alpha\!=\!0$ ($\lambda_1 \!=\! 0.362$), ($b$)~$\alpha\!=\!0.5$ ($\lambda_1 \!=\! 0.349$), ($c$)~$\alpha\!=\!1$ ($\lambda_1 \!=\! 0.375$). Remaining parameters: $r\!=\!2.8$, $\epsilon\!=\!0.7$, $N \!=\!256$.}
\label{fig:spacetime}
\end{figure}

The spatiotemporal chaos of a lattice of 256 coupled maps is shown in Fig.~\ref{fig:spacetime} for $r\!=\!2.8$, $\epsilon\!=\!0.7$, and three values of $\alpha$. Each lattice has been iterated forward in time for $1 \!\times\! 10^6$ time steps to allow initial transients to decay.  Figure~\ref{fig:spacetime} shows space-time diagrams of 512 alternating time steps after this initial warm up period using color contours of $u_i$. Alternating time steps are shown to highlight spatial structures present in the dynamics that appear as streaks persisting for many time steps.

\subsection{Numerical Stability}
\label{section:stability}

The dynamics of a map on the lattice can be thought of as a single map which is perturbed by the dynamics of its neighbors through the coupling terms in Eq.~\eqref{eq:cml}. Therefore, for values of $r$ that approach $r_c$ from below, it is possible for these fluctuations to push one of the maps on the lattice into the numerically unstable regime, $u_i \!<\! u_-$,  which could cause the lattice values to diverge toward negative infinity. The final stability outcome of the lattice depends upon specific contributions of the nearest neighbors in the subsequent time steps as we discuss in more detail below.

The numerical stability of the lattice is limited by the values of $\epsilon$ and $\alpha$ that are required to yield bounded dynamics. More insight can be gained by expressing Eq.~\eqref{eq:cml} as the weighted sum 
\begin{multline}
u_i^{(n+1)} = w_l^{(n)} f(u_{i-1}^{(n)}) + w_0 f(u_i^{(n)}) + w_r^{(n)} f(u_{i+1}^{(n)}) 
\label{eq:weights}
\end{multline}
with the weights $w_l^{(n)} = \frac{\epsilon}{2} + \frac{\alpha}{2} f(u_i^{(n)})$, $
w_0 = 1 - \epsilon$, and $w_r^{(n)} = \frac{\epsilon}{2} - \frac{\alpha}{2} f(u_i^{(n)})$.
We highlight that the weights on the left and right, $w_l^{(n)}$ and $w_r^{(n)}$, respectively, are time varying while $w_0$ is a constant. Importantly, the sum of the weights $w_l^{(n)} \!+\! w_0 \!+\! w_r^{(n)} \!=\! 1$ for all $n$.

By the maximum principle~\cite{patankar:1980,hundsdorfer:2003,horn:2012}, if the weights sum to unity and each weight is non-negative, the resulting value $u_i^{(n+1)}$ is guaranteed to be within the minimum and maximum values of $f(u^{(n)})$ that appear in the sum on the right hand side of Eq.~\eqref{eq:weights}. This can be expressed as $f(u^{(n)})|_{\text{min}} \!\le\! u_i^{(n+1)} \!\le\! f(u^{(n)})|_{\text{max}}$ where the minimum and maximum values are of $f(u_{i-1}^{(n)})$, $f(u_i^{(n)})$, and $f(u_{i+1}^{(n)})$. Therefore, the dynamics will remain numerically stable at the new time step if they were stable at the previous time step.

Conversely, if any of the weights are negative, the value $u_i^{(n+1)}$ may be outside of the minimum and maximum values of $f(u_{i-1}^{(n)})$, $f(u_i^{(n)})$, and $f(u_{i+1}^{(n)})$. As result, it is possible that $u_i^{(n+1)} \!<\! u_-$ which could cause the lattice to diverge to negative infinity.

The requirement that the weights are non-negative yields the following conditions for the numerical stability of the lattice, $\epsilon \!\le\! 1$ and $\alpha \!\le\! \frac{4 \epsilon}{r}$. The equalities of these expressions yield upper bounds for stable lattice dynamics. If either condition is violated, this indicates that a site on the lattice can have a value that is outside of the range of the values of the previous time step, it does not guarantee numerical instability. The numerical stability of the lattice depends upon the particular dynamics of the nearest neighbors to determine if the lattice site proceeds to diverge  or if it is pulled back within the stable range in the subsequent time steps.

\begin{figure}
\begin{center}
\includegraphics[width=2.5in]{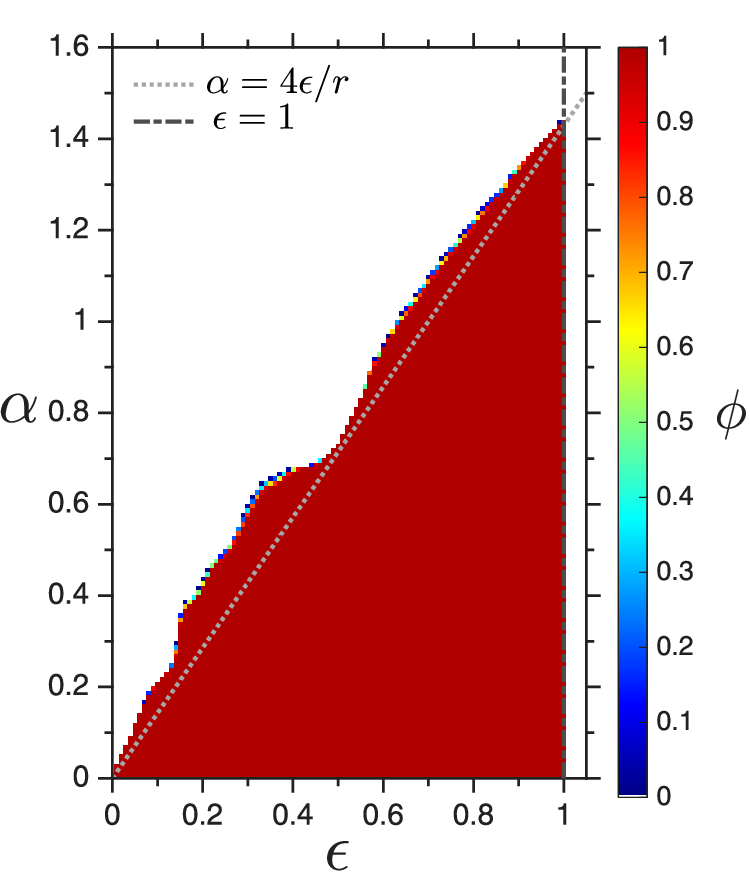}
\end{center}
\vspace{-9pt}
\caption{The fraction $\phi$ of 100 random initial conditions leading to numerically stable dynamics for a lattice with $r\!=\!2.8$, $N\!=\!256$, and $\Delta \alpha \!=\! \Delta \epsilon \!=\! 0.01$. Each simulation is run for $10^5$ time steps. Color contours of $\phi$ where red is a large fraction, blue is a small fraction, and white indicates $\phi \!=\!0$.  Random initial conditions are chosen where each lattice site is initially within bounds of the chaotic attractor for a single lattice. Numerical stability boundaries are indicated by the the vertical dashed line, $\epsilon \!=\!1$, and the diagonal dashed line, $\alpha \!=\! 4 \epsilon/r$.}
\label{fig:stability}
\end{figure}

We quantify this in Fig.~\ref{fig:stability} for a lattice with $r\!=\!2.8$ and $N\!=\!256$. For each possible pair ($\alpha$,$\epsilon$) we conduct 100 simulations initiated from randomly chosen initial conditions with a grid spacing of $\Delta \epsilon \!=\! \Delta \alpha \!=\! 0.01$. The initial conditions are chosen such that all lattice sites are within the bounds of the chaotic attractor for an isolated quadratic map, $f(r/4) \!\le\! u_i^{(1)} \!\le\! f(0)$ for all $i$.  Each simulation is evolved for $10^5$ time steps. Color contours represent the fraction $\phi$ of initial conditions which lead to numerically stable dynamics for the lattice. Dark red indicates $\phi \!=\!1$ where all initial conditions lead to numerically stable dynamics. White indicates $\phi \!=\! 0$ where zero of the initial conditions yielded numerically stable dynamics. As the color goes from red to blue the value of $\phi$ is decreasing. Figure~\ref{fig:stability} indicates numerical stability and it does not convey what kind of dynamics exist at those parameter values. The specific dynamics that have been realized are quantified in \S\ref{section:lyapunov-exponents} and \S\ref{section:fractal-dimension} using the Lyapunov exponents and fractal dimension, respectively.

The stability boundary $\epsilon \!=\!1$ is shown by the vertical dashed line. This is a sharp boundary, when $\epsilon \!>\! 1$ the weight $w_0$ is  negative for every evaluation of Eq.~\eqref{eq:weights}. This leads to many opportunities for the lattice dynamics to diverge toward negative infinity. For our sample of the dynamics shown Fig.~\ref{fig:stability}, all cases explored with $\epsilon \!>\! 1$ become numerically unstable by $10^5$ time steps for all of the initial conditions we explored.

The stability boundary $\alpha \!=\! 4 \epsilon/r$ is shown by the diagonal dashed. This boundary is not a sharp limit on the dynamics. For $\alpha \!\le\! 4 \epsilon/r$ all cases are numerically stable, $\phi \!=\!1$, as expected since it is not possible for a lattice site to diverge. However, we note that this condition is a result of requiring $w_l^{(n)}$ and $w_r^{(n)}$ to be non-negative. Therefore, for the lattice to become unstable requires a lattice site to become significantly negative while its nearest neighbors do not pull it back into the stable regime. As a result, this depends upon the specific lattice dynamics that are visited. This yields the interesting upper limit of numerically stable dynamics in Fig.~\ref{fig:stability} indicated by the non-white edge above $\alpha = 4 \epsilon/r$. The dynamics at the edge between the red and white regions is quite complex and is the only place on the stability diagram to yield values of $\phi$ that are neither zero or one. Overall, the numerical stability properties of the lattice limit our investigation to be within the non-white region of Fig.~\ref{fig:stability}.

\section{Results}
\label{section:results}

\subsection{Chaotic Dynamics with Drift}
\label{section:drift}

\begin{figure}
\begin{center}
\includegraphics[width=3.3in]{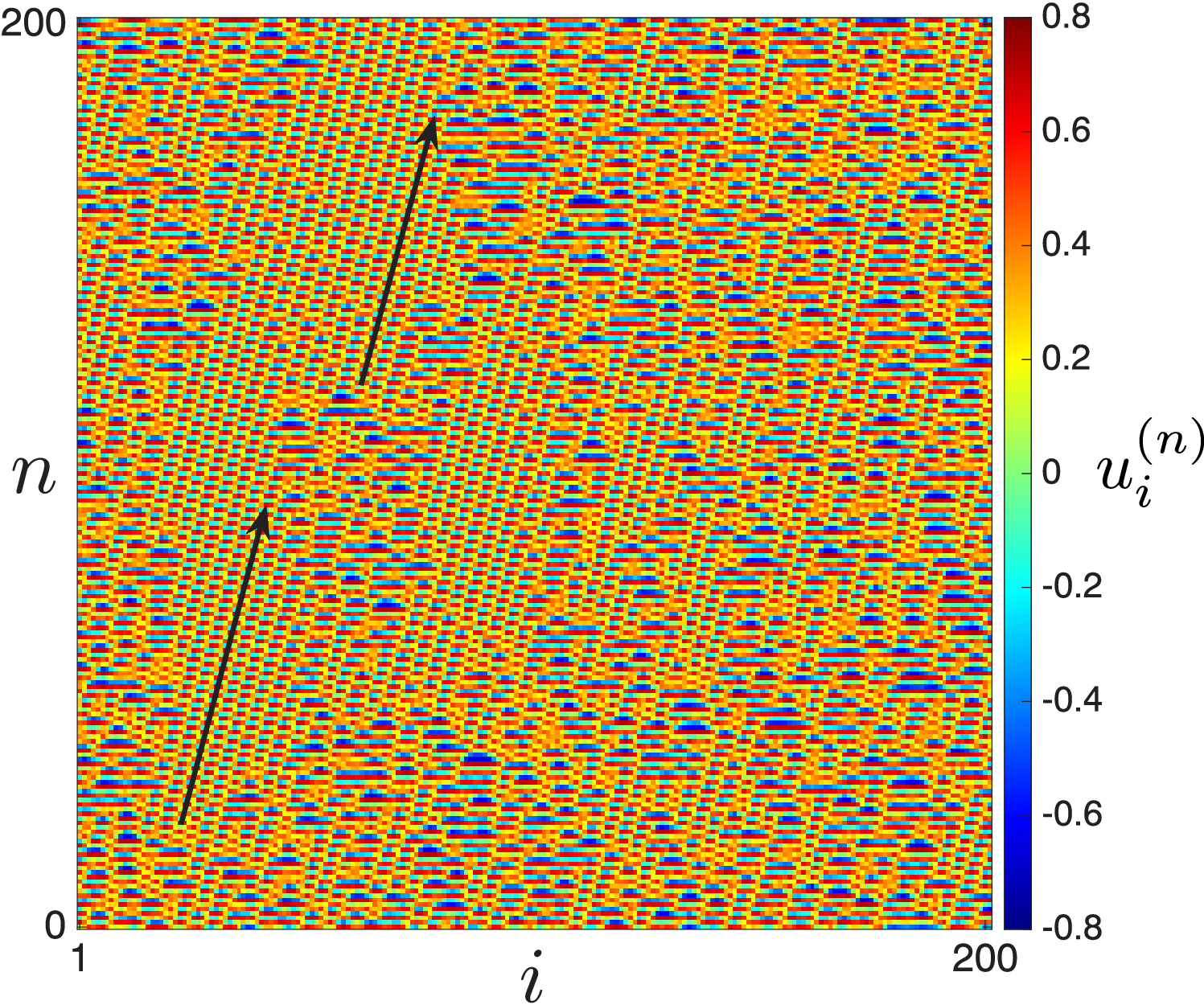}
\end{center}
\vspace{-9pt}
\caption{Spacetime plot of chaotic dynamics for $\alpha\!=\!0.5$.  Color contours of $u_i^{(n)}$ show $200$ time steps after initial transients have decayed. A zoomed-in view of $200$ lattice sites is shown to illustrate the details of the dynamics.  Two black arrows indicate the drift of spatial structures. The drift velocity $v_d$ is the inverse slope of the arrows: $v_d \!=\! 0.279$ (lower arrow), $v_d \!=\! 0.268$ (upper arrow). Remaining parameters: $r\!=\!2.8$, $\epsilon\!=\!0.7$, $N \!=\!256$.}
\label{fig:spacetime-closeup}
\end{figure}

In the absence of convective coupling, ($\alpha$=$0$, $\epsilon \!>\! 0$), the spatial structures in a spacetime plot of the dynamics align vertically as shown in Fig.~\ref{fig:spacetime}($a$). The presence of drifting spatial structures for $\alpha \!>\! 0$ is evident in Fig.~\ref{fig:spacetime}(b)-(c) where  the streak structures travel from left to right with increasing time. The amount of the drift increases with increasing $\alpha$. A close-up view of the dynamics for $\alpha \!=\! 0.5$ is shown in Fig.~\ref{fig:spacetime-closeup} for 200 time steps. A subset of $200$ lattice sites is shown to clearly display the details of the intricate dynamics. The drift of the spatial structures is indicated in two places by black arrows aligned with the streaks. The inverse slope of an arrow yields an estimate of the drift velocity $v_d$, for the arrows shown $v_d \!=\! (0.279,0.268)$.

\begin{figure}
\begin{center}
\includegraphics[width=3in]{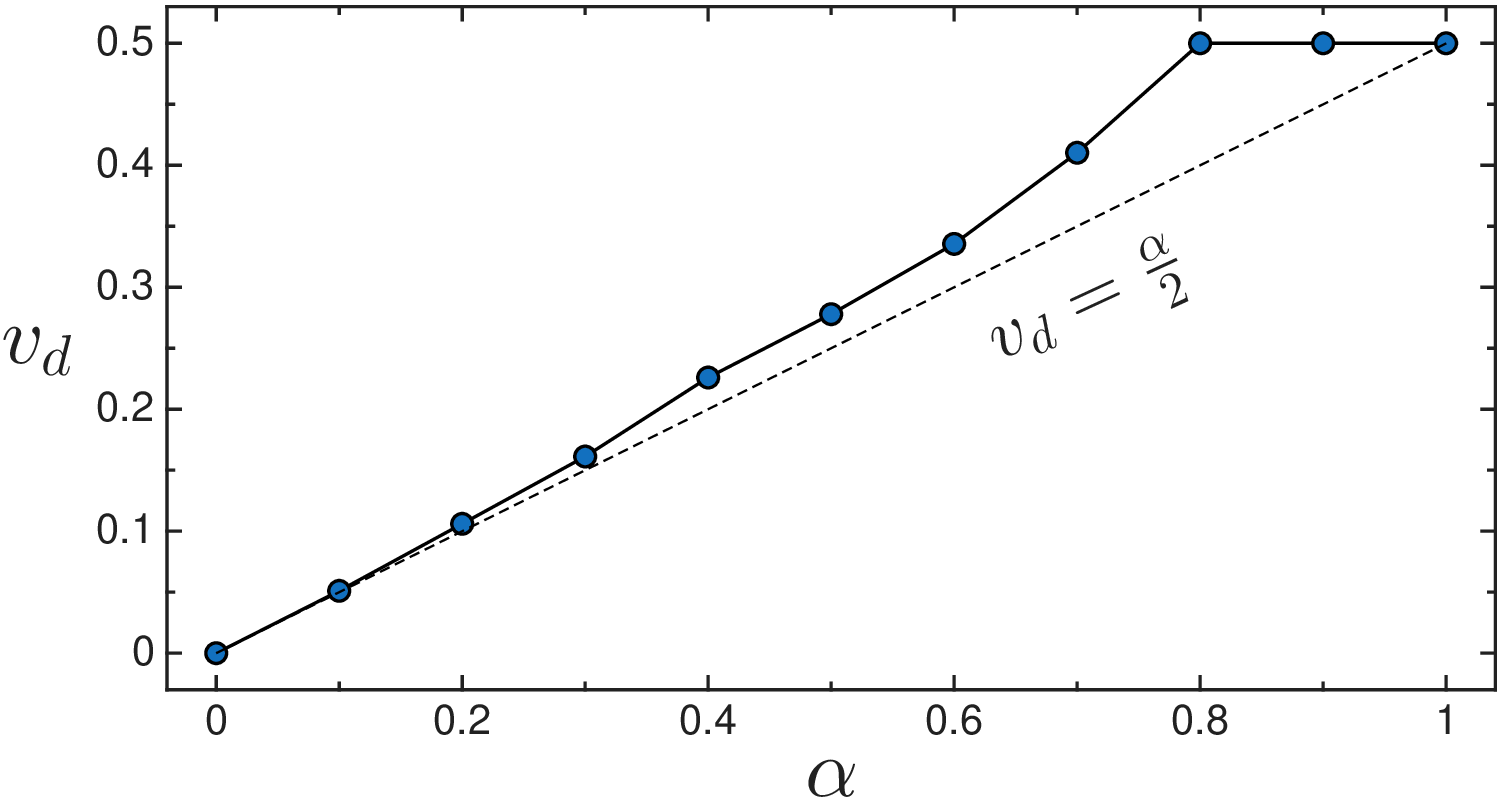}
\end{center}
\vspace{-9pt}
\caption{The variation of $v_d$ with $\alpha$. The dashed line is the linear trend $v_d \!=\! \alpha/2$. Remaining parameters: $r\!=\!2.8$, $\epsilon\!=\!0.7$, $N \!=\!256$.}
\label{fig:vd}
\end{figure}

The variation of $v_d$ with $\alpha$ is shown in Fig.~\ref{fig:vd}. Each symbol is the average of the drift velocity using 5 streaks.  Overall, the drift velocity increases approximately linearly with $\alpha$ where $v_d \!=\! \alpha/2$ is shown by the dashed line. For larger values of convective coupling, $\alpha \gtrsim 0.5$, $v_d$ increases faster than the linear trend and reaches a maximum of $v_d \!=\! 0.5$ for $\alpha \ge 0.8$. 

The positive value of $v_d$ is due to the convective coupling and our choice of mapping function given by Eq.~\eqref{eq:quadratic-map}. At any time $n$, the spatial average of $f(u)$ on the lattice is a positive number. This can be expressed as $\langle f(u_i^{(n)})\rangle_i \!>\! 0$ where the angle brackets indicate an average over the lattice sites $i$. This quantity acts like a constant wave speed. For example, for $\alpha \!=\! 0.5$, if we average over space and time the result is $\langle f(u_i)^{(n)} \rangle_{i,n}\!=\!0.232$ which yields a value that is approximately equal to the drift velocity.

\begin{figure}
\begin{center}
\includegraphics[width=2.4in]{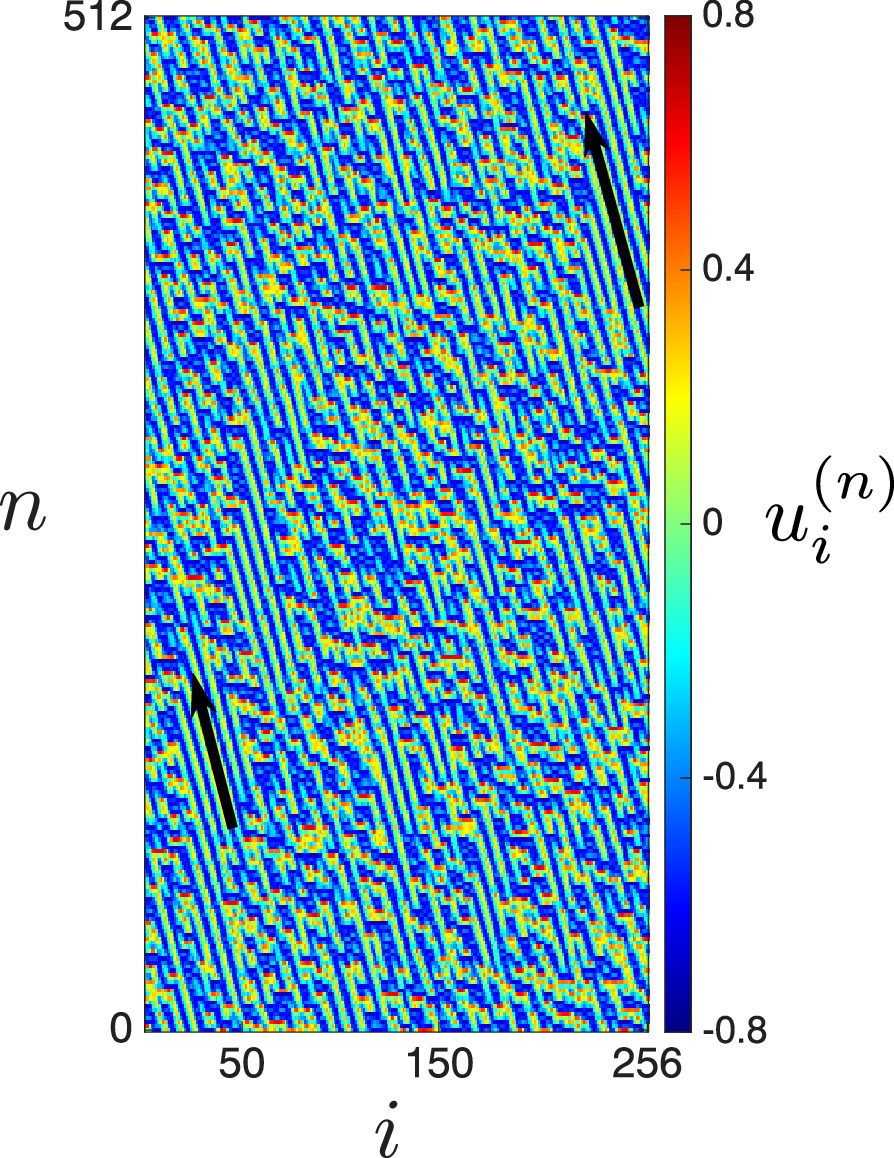}
\end{center}
\vspace{-9pt}
\caption{Spacetime plot of $u_i^{(n)}$ for a lattice with  $\alpha\!=\!0.5$ where $f(u)$ has been replaced with $-f(u)$ in  Eq.~\eqref{eq:cml}. This exhibits a drift of spatial structures from right to left with increasing time. The data is plotted every other time step. Remaining parameters: $r\!=\!2.8$, $\epsilon\!=\!0.7$, $N \!=\!256$.}
\label{fig:spacetime-reverse}
\end{figure}

As a result of this dependence of $v_d$ on the $f(u)$, the direction of the drift velocity can be reversed by replacing $f(u)$ with $-f(u)$ in Eq.~\eqref{eq:cml} to  reverse this constant wave speed. This yields the spacetime dynamics shown in Fig.~\ref{fig:spacetime-reverse} where the drift velocity is now negative with the drift going from right to left.

\subsection{Lyapunov Vectors and Exponents}
\label{section:lyapunov-exponents}

The CLVs provide important insights into the chaotic dynamics by quantifying the directions of growth or decay of small perturbations in the tangent space. For a lattice with $N$ sites, there are $N$ CLVs, $\vec{v}_k$, where $k=1,2,\ldots,N$ is the Lyapunov index. Every CLV is a $N$-dimensional vector that grows, or decays, exponentially with a rate given by its Lyapunov exponent $\lambda_k$.  We use the dynamic algorithm~\cite{ginelli:2007} to calculate the CLVs, a detailed discussion of the CLV calculation using our conventions is given in Ref.~\cite{raj:2024}.

The dynamics of the perturbations are given by the tangent space equations which are the  linearization of Eq.~\eqref{eq:cml}. First, the lattice dynamics are integrated forward in time for $2\!\times\!10^6$ time steps to allow all initial transients to decay.  $N$ copies of the tangent-space equations are then simultaneously evolved forward in time, along with Eq.~\eqref{eq:cml}, for $5\!\times\!10^4$ time steps to determine the dynamics of the $N$ perturbation vectors $\delta \vec{u}_k^{(n)}$. During this computation, periodic $\mathbf{QR}$ decompositions are performed on the matrix containing the perturbation vectors as its columns, both the $\mathbf{Q}$ and $\mathbf{R}$ matrices are stored for later use. The columns of $\mathbf{Q}$ contain the orthogonal (Gram-Schmidt) Lyapunov vectors that are essential for the computation of CLVs.

The tangent-space equations are then evolved backwards in time for $\!\gtrsim\!5\!\times\!10^3$ time steps while the converged CLVs are determined as a linear combination of the orthogonal Lyapunov vectors. At this point, a set of converged CLVs are computed while continuing to evolve the equations backwards in time for an additional $\!4\!\times\!10^4$ time steps. Each of the converged CLVs are then  evolved forward in time using the tangent space equations to compute values of the finite-time covariant Lyapunov exponents (CLEs), $\tilde{\lambda}_k$. The long-time average of $\tilde{\lambda}_k$ converges to the Lyapunov exponents $\lambda_k$ where $\lambda_1 \!\ge\! \lambda_2 \ldots \!\ge\! \lambda_N$ and $\lambda_1 \!>\! 0$ indicates chaos. 

\begin{figure}
\begin{center}
\includegraphics[width=2.5in]{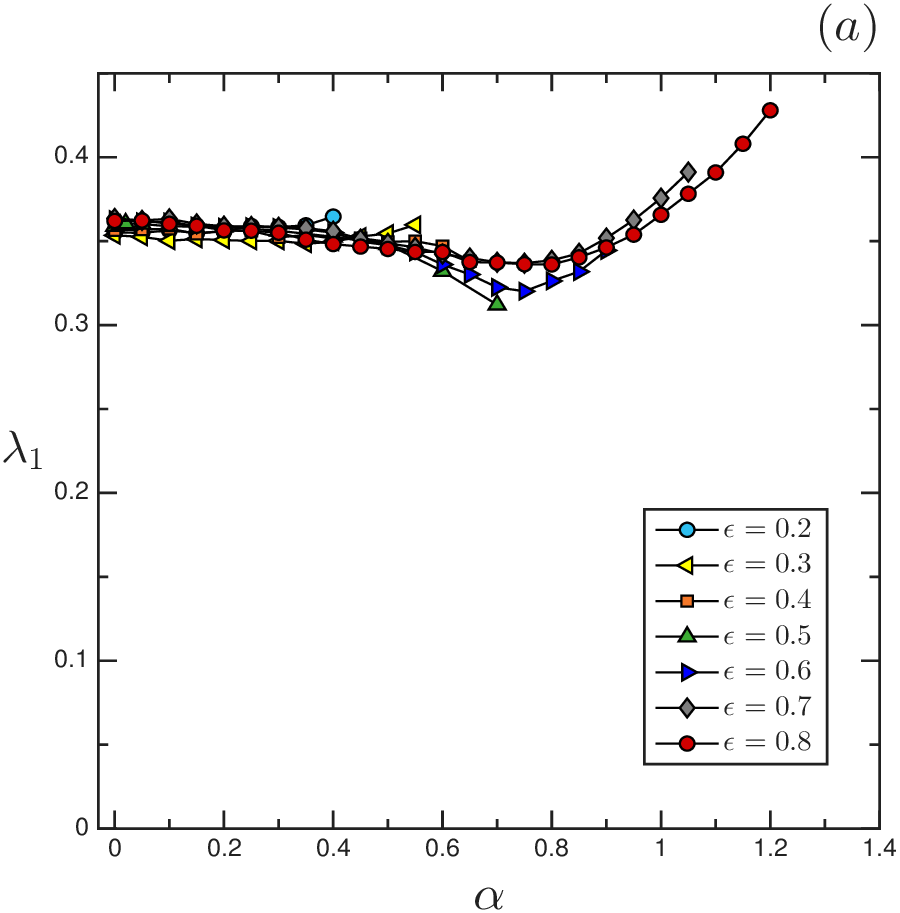} \\ 
\includegraphics[width=2.5in]{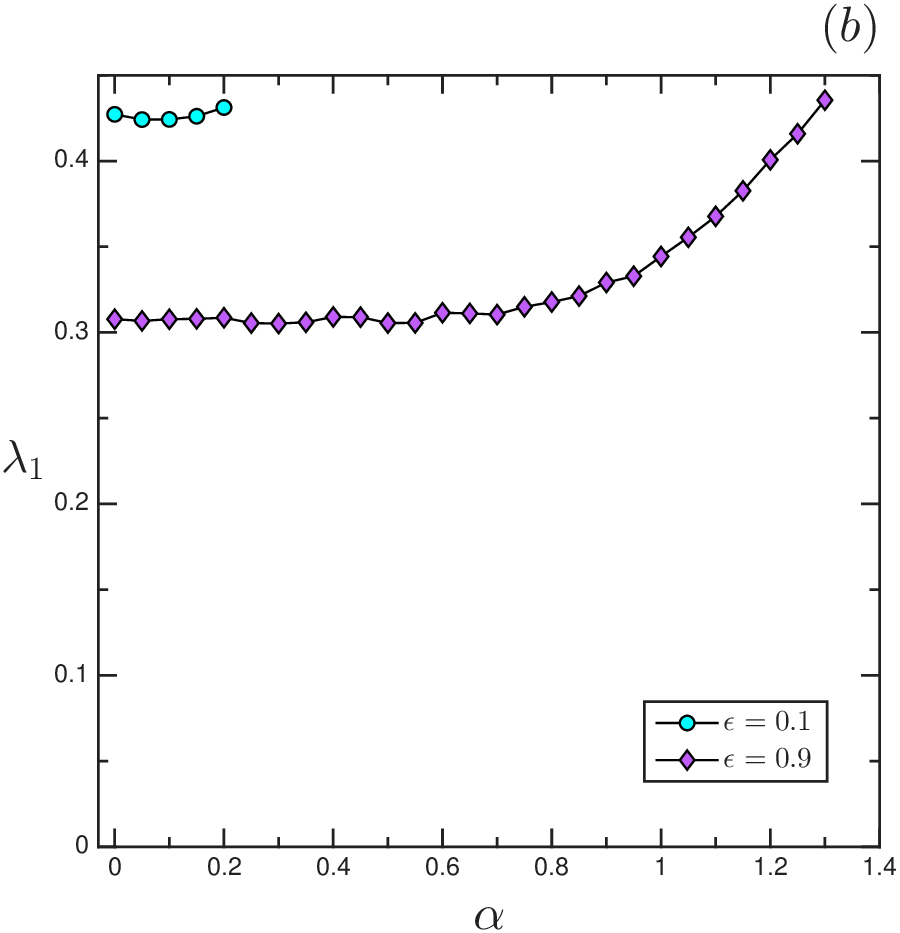}
\end{center}
\vspace{-9pt}
\caption{The variation of the leading Lyapunov exponent, $\lambda_1$, with the strength of convection, $\alpha$, and diffusion, $\epsilon$. Results for ($a$) $0.2 \! \le \! \epsilon \!\le\! 0.8$ and ($b$) $\epsilon \!=\! 0.1$ and $\epsilon \!=\! 0.9$. Remaining Parameters: $r\!=\!2.8$, $N\!=\!256$.}
\label{fig:lambda1-alpha}
\end{figure}

The variation of $\lambda_1$ with ($\alpha$,$\epsilon$) is shown in Fig.~\ref{fig:lambda1-alpha}. For $0.2 \!\lesssim\! \epsilon \!\lesssim\! 0.8$ the leading Lyapunov exponent does not vary significantly with $\epsilon$ as shown in Fig.~\ref{fig:lambda1-alpha}($a$).  However, for small and large values of $\epsilon$, the values of $\lambda_1$ are different than this general trend as shown in Fig.~\ref{fig:lambda1-alpha}($b$). For small amounts of diffusion, $\lambda_1$ increases to a value of $\lambda_1 \approx 0.43$ over the limited range of $\alpha$ that yields numerically stable dynamics. The lack of diffusion in the lattice yields a value of $\lambda_1$ that is closer to the value that would be found for an isolated map ($\lambda_0 \!=\! 0.6034$). For large values of $\epsilon$, the increased diffusion causes a significant decrease in $\lambda_1$ for small $\alpha$ which increases with increasing $\alpha$.

\begin{figure}
\begin{center}
\includegraphics[width=2.5in]{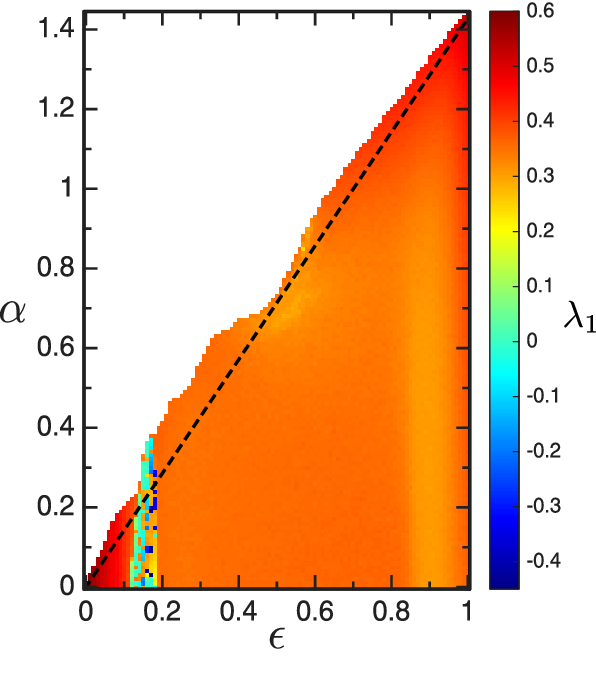}
\end{center}
\vspace{-9pt}
\caption{Variation of $\lambda_1$ with $\alpha$ and $\epsilon$ for a lattice with $r\!=\!2.8$, $N \!=\!256$ and $\Delta \alpha \!=\! \Delta \epsilon \!=\! 0.01$. White indicates the absence of numerically stable dynamics. The value of $\lambda_1$ for the numerically stable dynamics is indicated using the color bar. Each simulation is initiated using lattice values within the limits of the chaotic dynamics of an isolated lattice.}
\label{fig:lambda1-grid}
\end{figure}

We show the variation of $\lambda_1$ over the entire range of $\alpha$ and $\epsilon$ that yield numerically stable dynamics in Fig.~\ref{fig:lambda1-grid} for a lattice with $r\!=\!2.8$ and $N\!=\!256$. For each $(\alpha,\epsilon)$ pair, we conduct a numerical simulation initiated from random initial conditions that is allowed to evolve $10^5$ time steps. The leading Lyapunov exponent is calculated and is represented using the color scale shown. The dynamics are chaotic for nearly all parameter values except for a small band in the range $0.1 \!\lesssim\! \epsilon \!\lesssim\! 0.2$ which contains a complex collection of steady, periodic, and chaotic dynamics.  

\begin{figure}
\begin{center}
\includegraphics[width=2.4in]{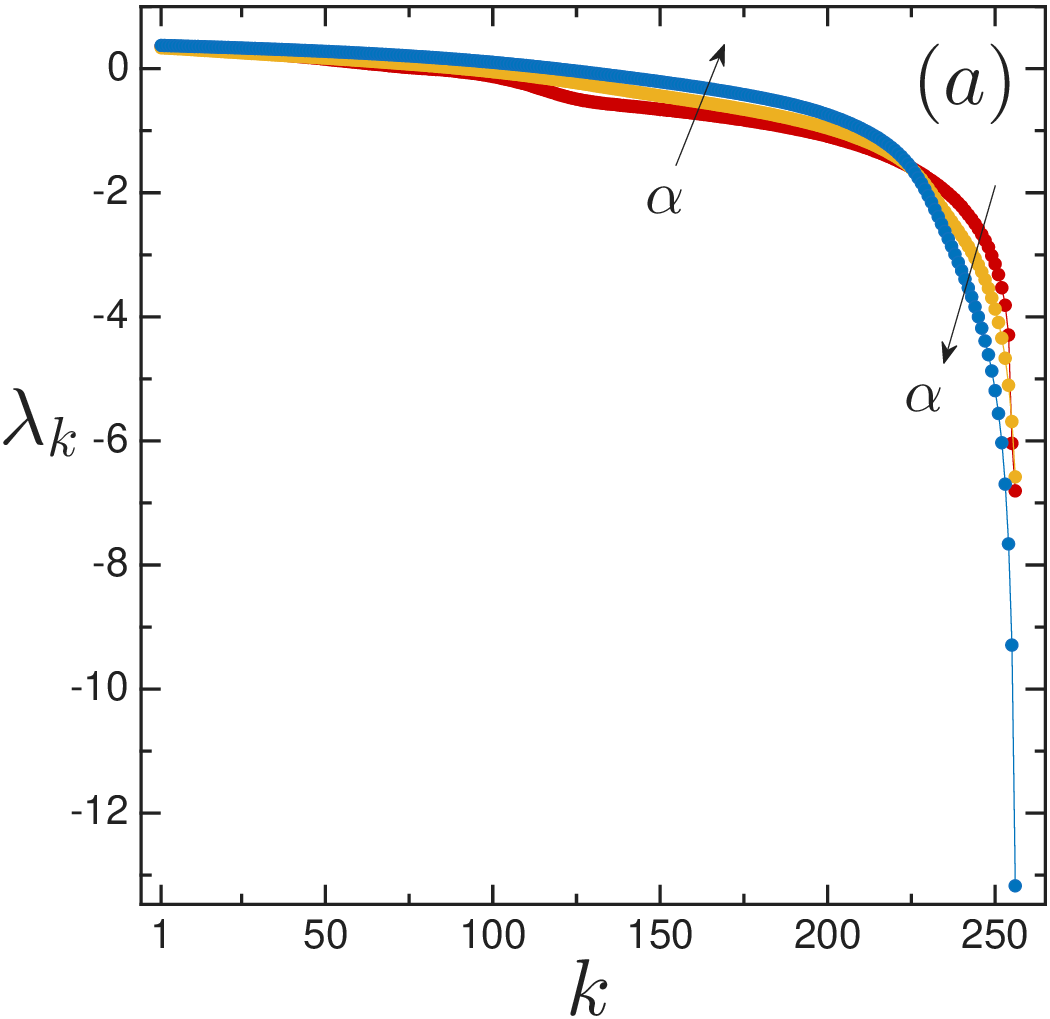} \hspace{1cm}
\includegraphics[width=2.4in]{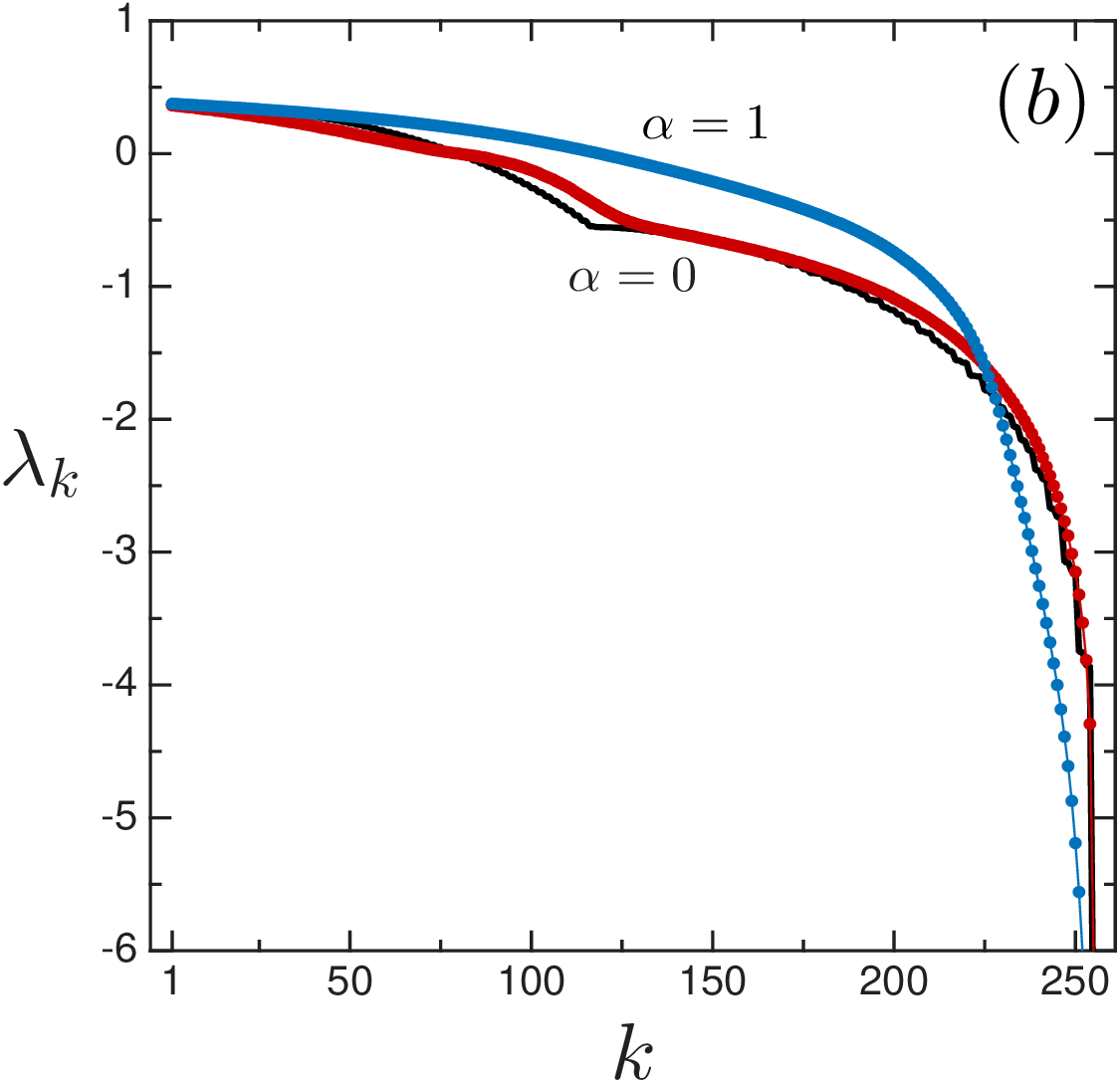}
\end{center}
\vspace{-9pt}
\caption{Variation of $\lambda_k$ with $\alpha$. ($a$)~$\alpha\!=\!0$ (red); $\alpha\!=\!0.7$ (orange); and $\alpha\!=\!1$ (blue). Arrows indicate trend for increasing $\alpha$. ($b$) Close-up view for $\alpha\!=\!0$ and $\alpha\!=\!1$. The black line is the theoretical prediction for $\alpha \!=\! 0$. Remaining parameters: $r \!=\!2.8$, $\epsilon\!=\!0.7$, $N\!=\!256$. }
\label{fig:lyapunov-spectra}
\end{figure}

The spectrum of Lyapunov exponents $\lambda_k$ is shown in  Fig.~\ref{fig:lyapunov-spectra}($a$). The arrows indicate the direction in which the spectrum changes with increasing $\alpha$. As $\alpha$ is increased, the Lyapunov exponents for $k \!\lesssim \!225$ increase and for $k \!\gtrsim \!225$ the Lyapunov exponents decrease. For small values of the convective coupling strength, $0\!\le\!\alpha \!\lesssim \!0.4$, the convective coupling does not significantly affect the Lyapunov exponents.

It is insightful to carefully explore the contributions of the spatial coupling on the variation of $\lambda_k$. We first express the time evolution of the lattice, given by Eq.~\eqref{eq:cml}, as
\begin{equation}
    \vec{u}^{(n+1)}=\mathbf{A}_c\vec{f}(\vec{u}^{(n)})+\mathcal{N}\!\left(\vec{f}(\vec{u}^{(n)})\right)
    \label{eq:operators}
\end{equation}
where $\vec{u}^{(n)}$ is a vector with $N$ components representing the state of the lattice at time $n$. The $N \!\!\times\!\! N$ matrix describing the diffusive spatial coupling is  $\mathbf{A}_c$ and $\mathcal{N}$ is the nonlinear operator that describes the convective coupling.

The eigendecomposition of the coupling matrix is $\mathbf{A}_c\vec{w}_j \!=\! \Lambda_j \vec{w}_j$ 
where $\Lambda_j$ and $\vec{w}_j$ are the eigenvalues and eigenvectors, respectively, and $j$ is the eigenindex $j \!=\! 1, 2, \ldots, N$.  For CMLs with diffusive coupling alone ($\epsilon \!>\!0$, $\alpha \!= \! 0$), an estimate of the Lyapunov exponents can be determined from the eigenvalues~\cite{raj:2024,takeuchi:2011}.

Figure~\ref{fig:lyapunov-spectra}(b) shows a close-up of the Lyapunov spectra for diffusive coupling alone, $\alpha\!=\!0$ (red), and for strong convective coupling, $\alpha\!=\! 1$ (blue). The black line is the theoretical prediction of $\lambda_k$ for $\alpha=0$ obtained using $\lambda_k = \lambda_1 + \ln|\Lambda_k|$ where $\lambda_1$ must be determined numerically. The eigenvalues have been sorted such that the $|\Lambda_k|$ are in descending order with the index $k$. A more accessible estimate can be found by setting $\lambda_1$ to equal the Lyapunov exponent of an isolated map, however this will typically overestimate the value of $\lambda_1$ that occurs for the lattice while correctly representing the shape of the spectrum~\cite{takeuchi:2011,raj:2024}.

The Lyapunov spectrum for $\alpha\!=\!0$ (red circles) closely follows the theoretical prediction as shown in Fig.~\ref{fig:lyapunov-spectra}($b$). For strong convective coupling, $\alpha\!=\!1$, the Lyapunov exponents deviate from this prediction and do not exhibit the elbow (kink) structure present for the diffusive alone case at $k \!\approx\! 125$. For $\alpha \!=\! 0$, the elbow in the spectrum occurs for $k$ corresponding to null eigenvalues of $\mathbf{A}_c$~\cite{raj:2024}.

\subsection{The Dimension of the Chaotic Dynamics}
\label{section:fractal-dimension}

The fractal dimension of the attractor describing the dynamics in state-space can be described using
\begin{equation}
D_\lambda = q + \frac{\sum_{k=1}^q \lambda_k}{|\lambda_{q+1}|}
\label{eq:fractal-dimension}
\end{equation}
where $q$ is the largest value of the index that yields a positive summation~\cite{kaplan:1979}. The fractal dimension represents the number of active chaotic degrees in the system, on average~\cite{farmer:1983}.  The value of $D_{\lambda}$ depends upon the shape of the spectrum of Lyapunov exponents and, as a result, it is sensitive to the spatial couplings that are present.

\begin{figure}
\begin{center}
\includegraphics[width=2.5in]{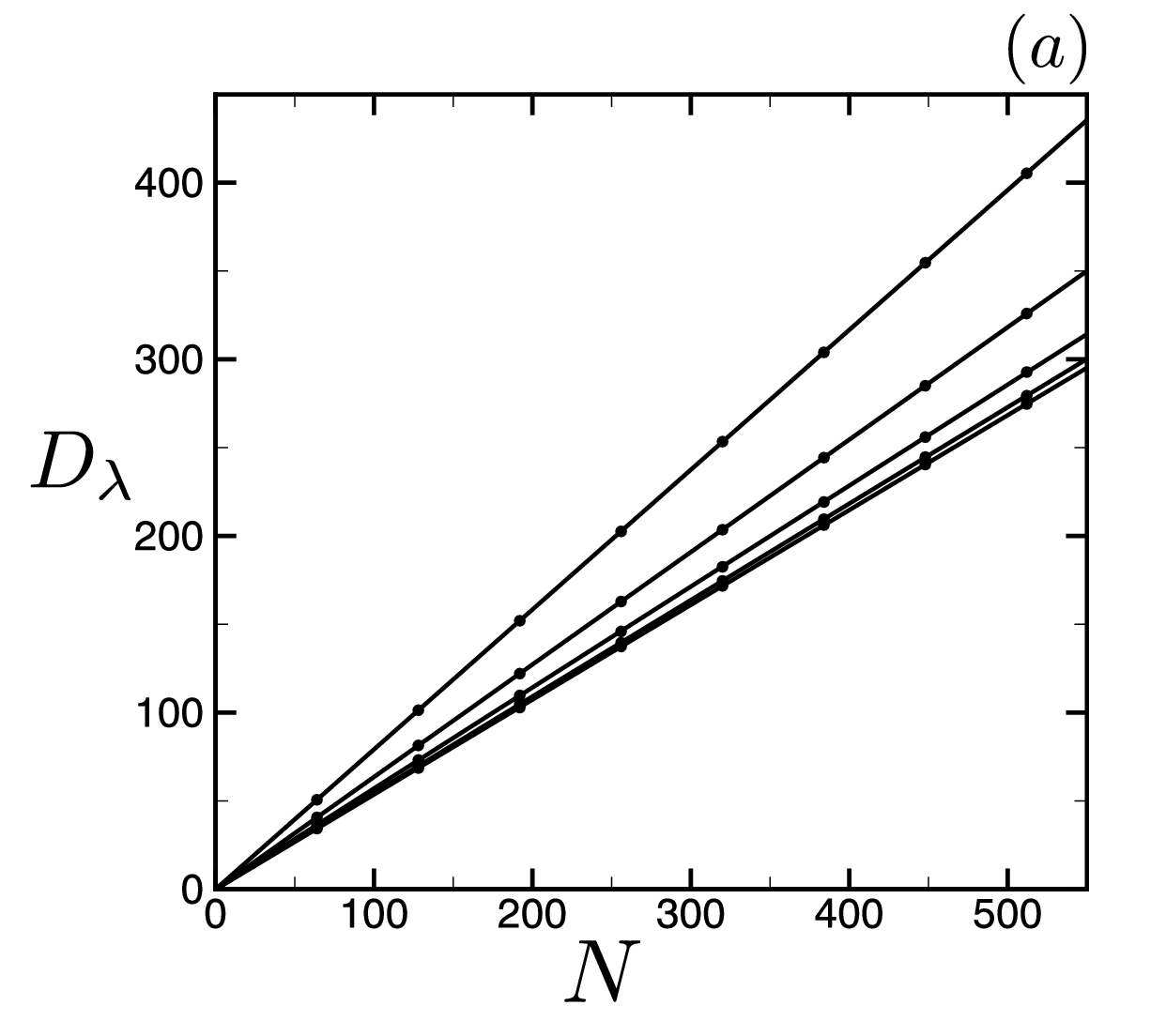} \\ 
\hspace{0.1cm} \includegraphics[width=2.5in]{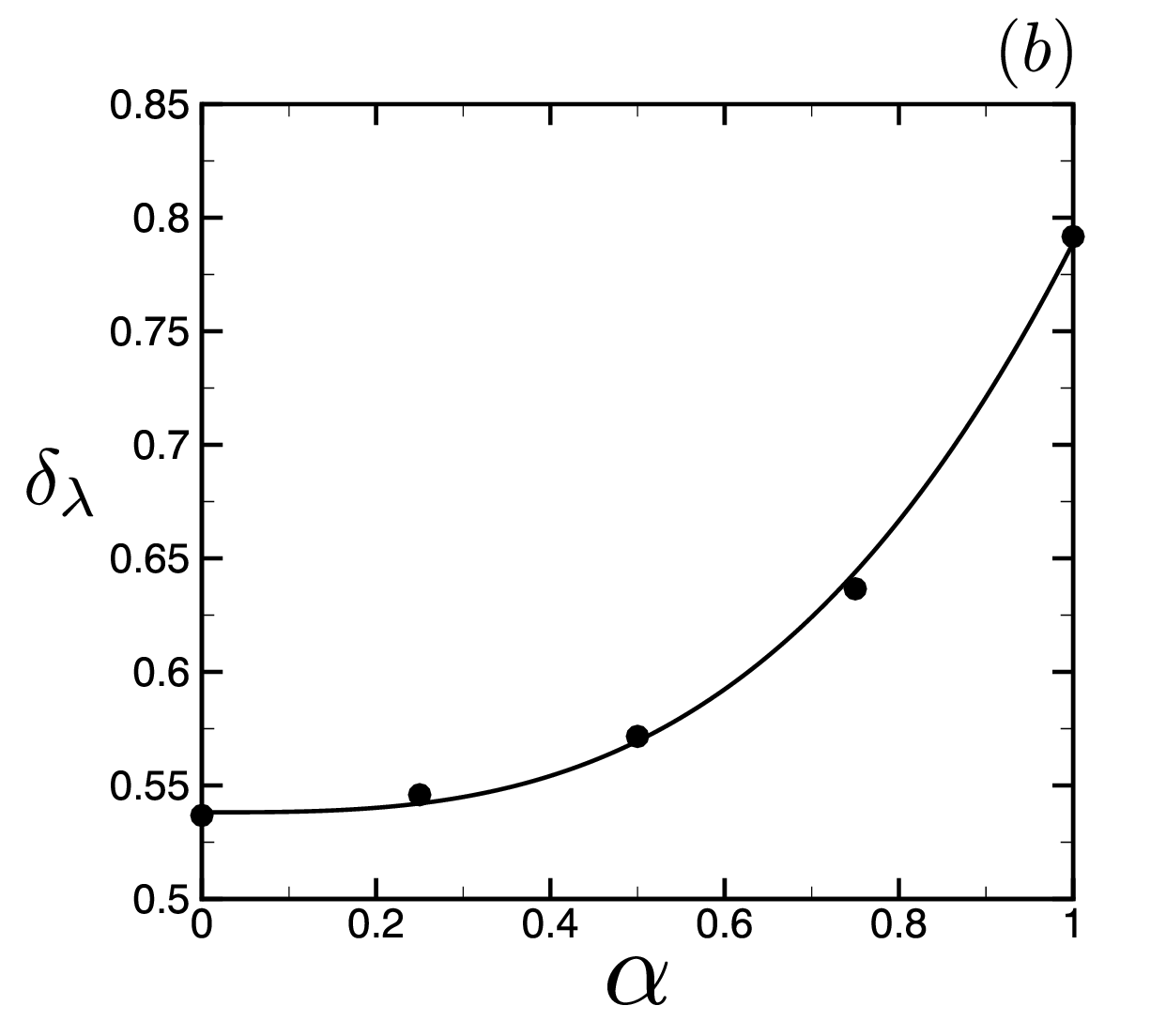} 
\end{center}
\vspace{-9pt}
\caption{Extensively chaotic dynamics. ($a$) $D_\lambda$ scales linearly with the system size $N$.  Solid lines are the fits $D_\lambda = \delta_\lambda N + D_0$ where $\delta_\lambda$ is the dimension density and $D_0$ is a fitting constant. Lines for each $\alpha = (0, 0.25, 0.5, 0.75, 1)$ occur in sequence from the lower to upper curve. ($b$) Dimension density as a function of $\alpha$, solid line is  $\delta_\lambda \!=\! 0.54 \!+\! 0.25 \alpha^3$. Remaining parameters: $r\!=\!2.8$, $\epsilon \!=\! 0.7$.}
\label{fig:extensivity}
\end{figure}

The variation of $D_\lambda$ with system size $N$ is shown in Fig.~\ref{fig:extensivity}($a$). Each curve is for a different value of $\alpha$ over the range $0 \!\le\! \alpha \!\le\! 1$. The lowest curve is for $\alpha=0$ and the remaining curves appear in sequence for increasing values of $\alpha$. The straight lines are fits $D_\lambda = \delta_\lambda N + D_0$ where $\delta_\lambda$ is the dimension density and $D_0$ is a constant.

A fractal dimension that is proportional to the size of the system, $D_\lambda \!\propto\! N$,  indicates extensive chaos~\cite{ruelle:1982}.  Figure~\ref{fig:extensivity}($a$) indicates that the results shown are in the extensively chaotic regime. In addition, the fractal dimension increases with increasing $\alpha$ resulting in increasing values of the dimension density $\delta_\lambda$ as shown in Fig.~\ref{fig:extensivity}($b$). The solid line is a cubic fit through the data.

\begin{figure}
\begin{center}
\includegraphics[width=2.5in]{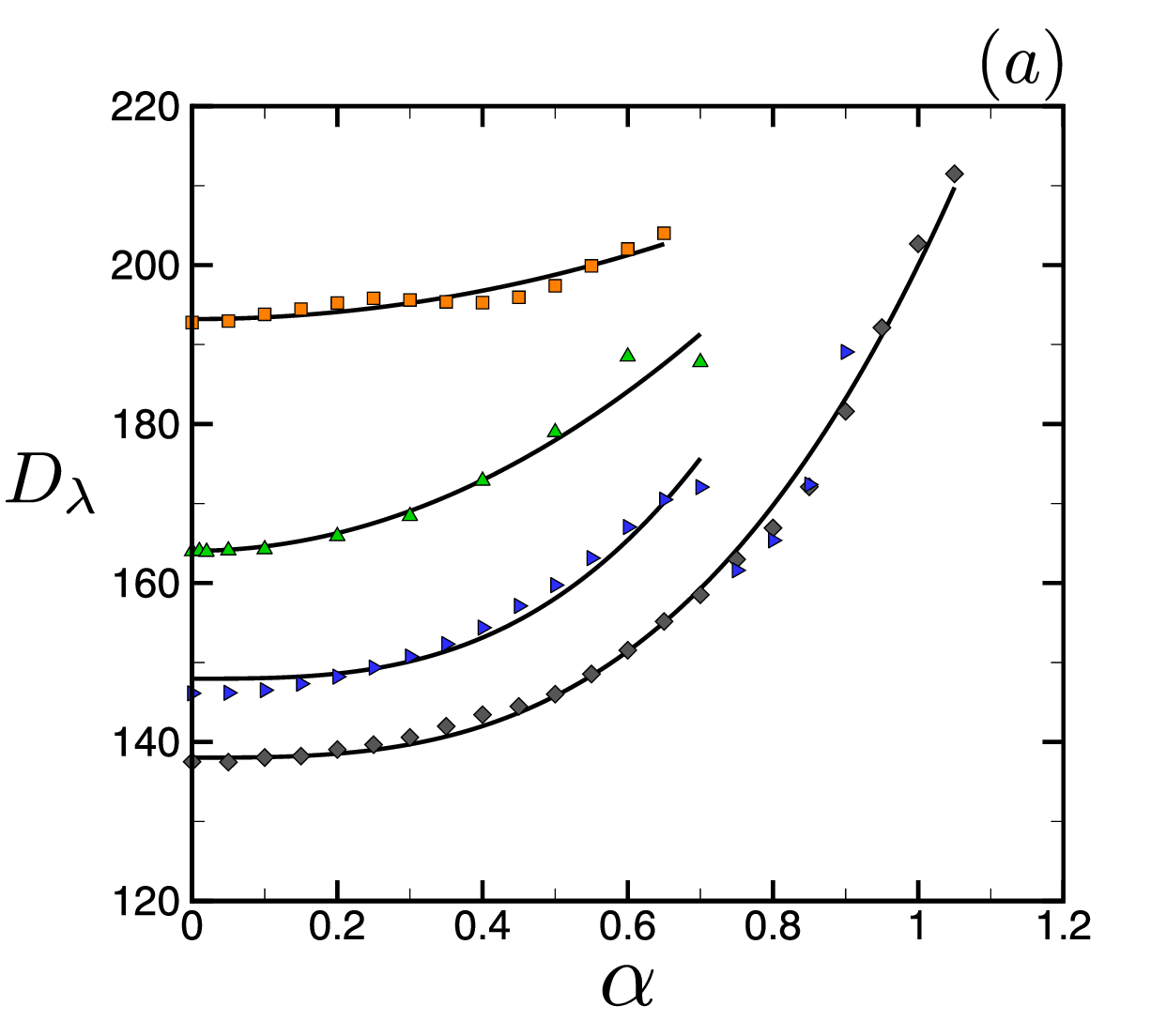}
\includegraphics[width=2.5in]{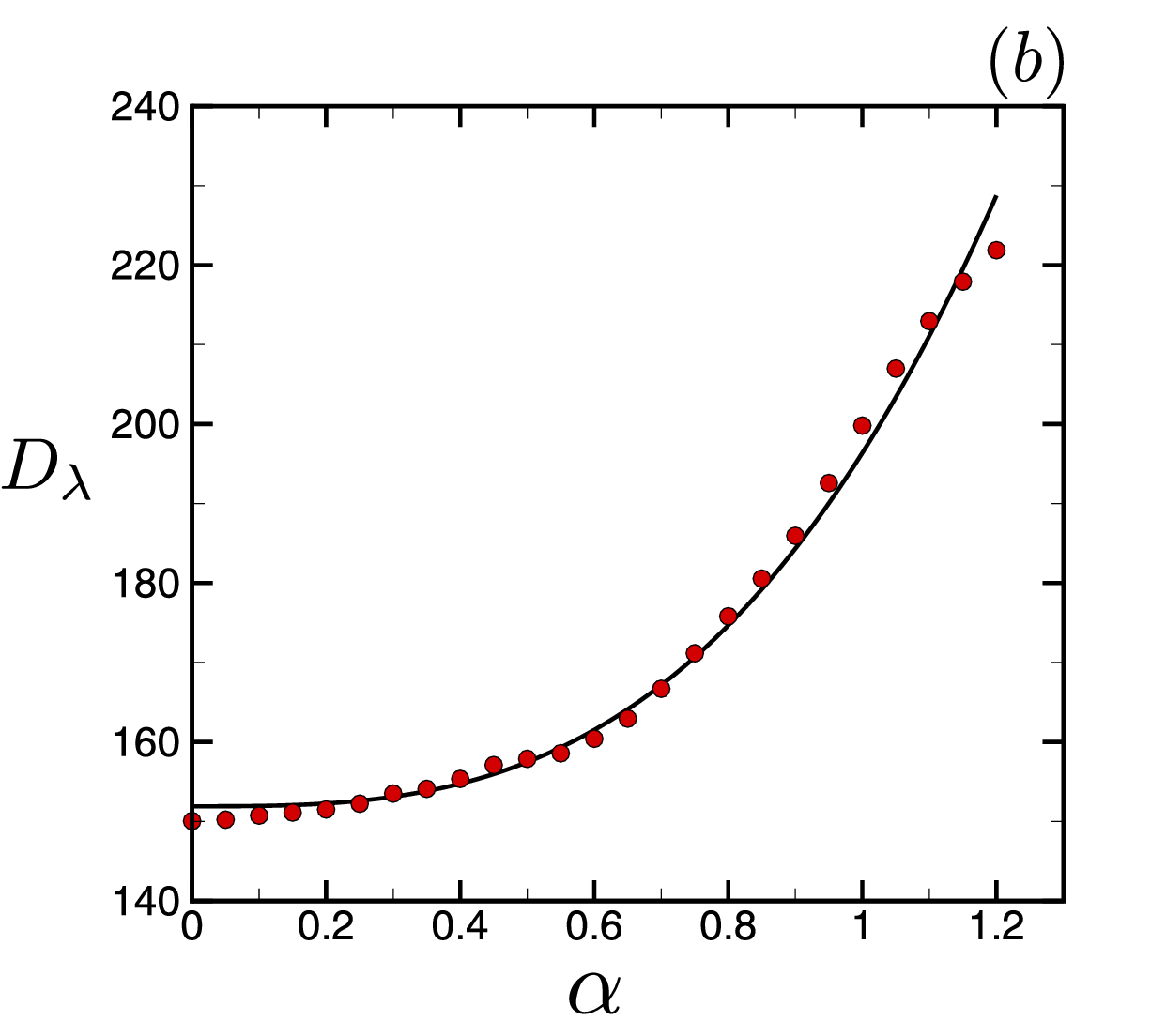}
\end{center}
\vspace{-9pt}
\caption{The variation of $D_{\lambda}$ with ($\alpha$, $\epsilon$). ($a$) $\epsilon \!=\! 0.4$ (orange, squares) with fit $D_\lambda \!=\! 193.2 \!+\! 22.3 \alpha^2$; $\epsilon \!=\! 0.5$ (green, triangles) with fit $D_\lambda \!=\! 164.0 \!+\! 55.7 \alpha^2$; $\epsilon \!=\! 0.6$ (blue, triangles) with fit $D_\lambda \!=\! 148.0 \!+\! 80.8 \alpha^3$; $\epsilon \!=\! 0.7$ (gray, diamonds) with fit $D_\lambda \!=\! 138.0 \!+\! 62.0 \alpha^3$. ($b$) $\epsilon \!=\! 0.8$ (red, circles) with fit $D_\lambda \!=\! 151.9 \!+\! 44.5 \alpha^3$.
Remaining parameters: $r\!=\!2.8$, $N \!=\!256$.}
\label{fig:fractal-dimension}
\end{figure}

Figure~\ref{fig:fractal-dimension} shows the rapid increase of $D_{\lambda}$ with respect to $\alpha$ for a lattice with $N \!=\! 256$. The solid lines are fits through the data. For all  conditions we have explored, we find $D_{\lambda}\!< N$. The increasing trend of $D_\lambda$ with $\alpha$ can be traced to the increase in  $\lambda_k$ with $k$ for $k \!\lesssim\! 225$ as shown in Fig.~\ref{fig:lyapunov-spectra}. It is important to highlight that this includes both positive and negative Lyapunov exponents.

\begin{figure}
\begin{center}
\includegraphics[width=2.5in]{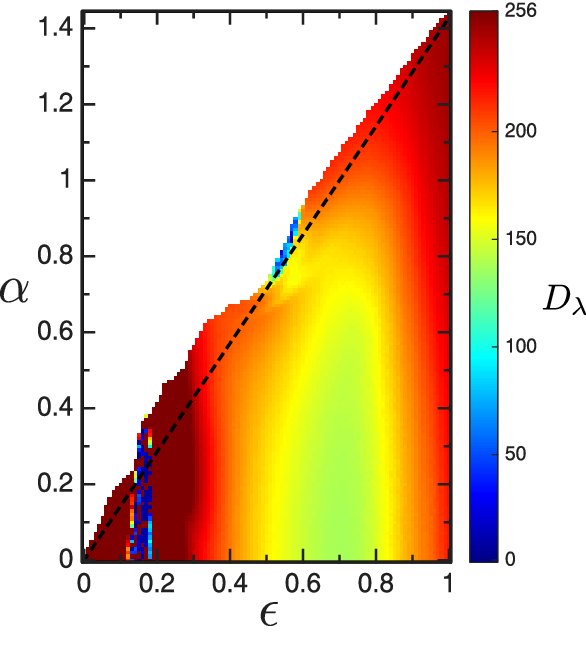}
\end{center}
\vspace{-9pt}
\caption{($a$) The variation of $D_\lambda$ with ($\alpha$, $\epsilon$) for $r\!=\!2.8$ and $N \!=\!256$ using the conventions of Fig.~\ref{fig:lambda1-grid}.}
\label{fig:fractal-grid}
\end{figure}

The variation of $D_\lambda$ over the entire range of ($\alpha$, $\epsilon$) is shown in Fig.~\ref{fig:fractal-grid} for $N=256$ and $r=2.8$. For each pair $(\alpha,\epsilon)$, the spectrum of Lyapunov exponents and fractal dimension are calculated.  The fractal dimension illustrates some interesting variations. Overall, $D_\lambda$ is quite large, $D\gtrsim 120$, for all parameter values except for the small band of at lower $\epsilon$ and an edge region at $\epsilon \approx 0.6$.

\subsection{The spatial structure of the CLVs}
\label{section:clvs}

\begin{figure}
\begin{center}
\includegraphics[width=1in]{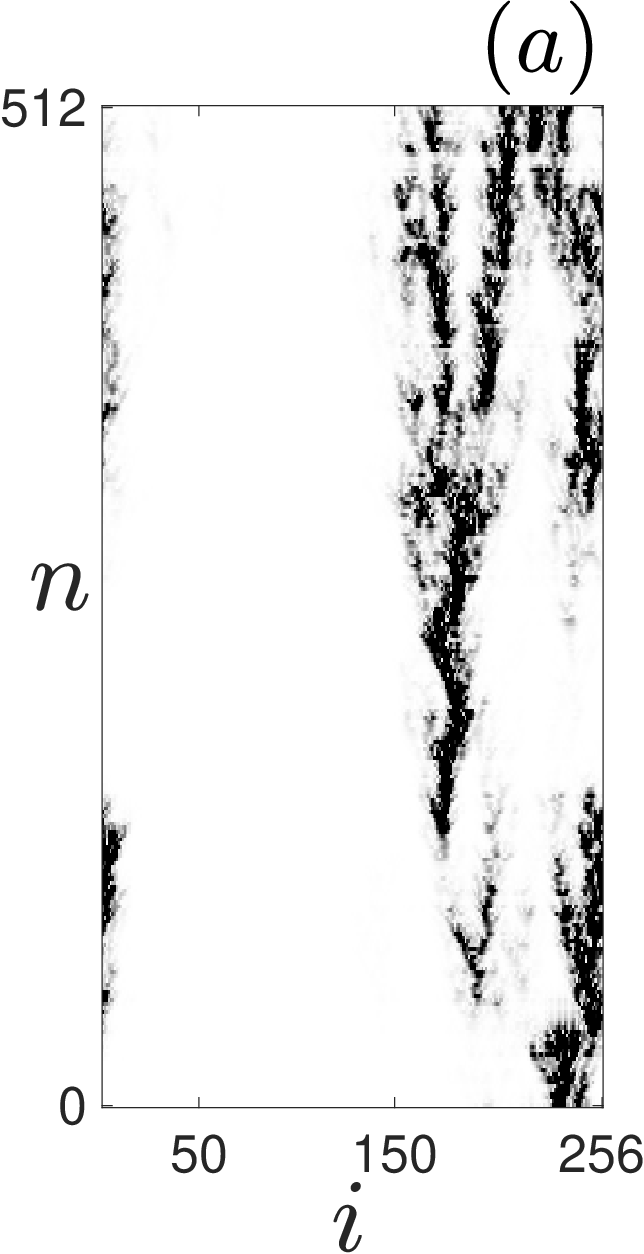} \hspace{0.1cm}
\includegraphics[width=1in]{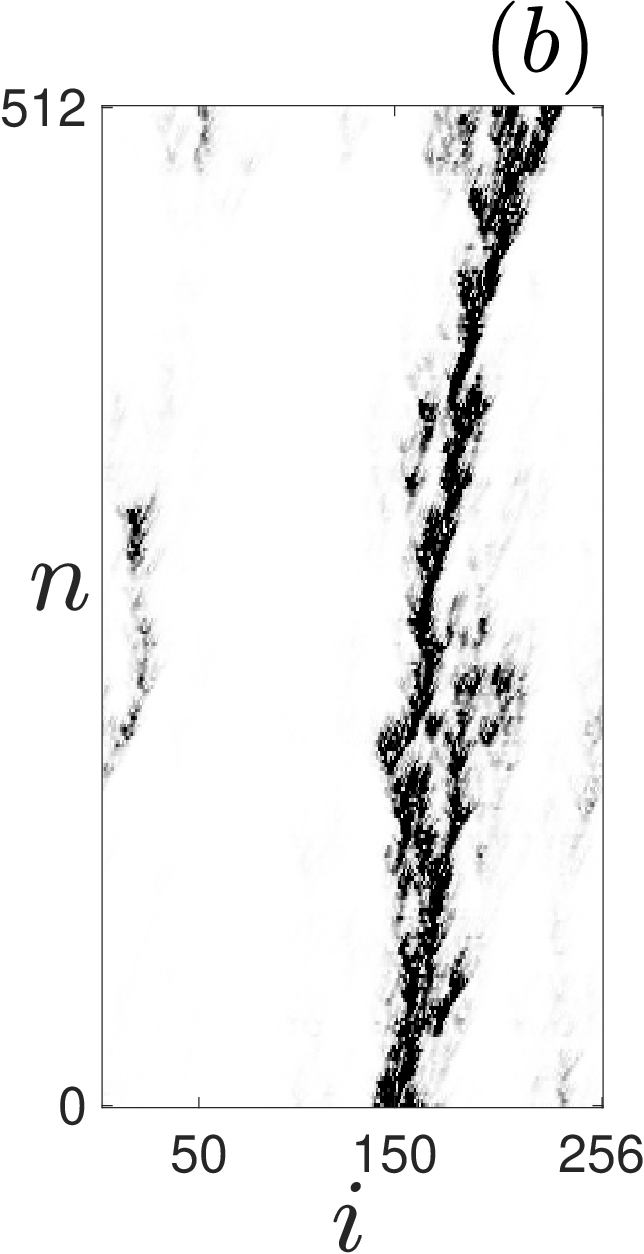} \hspace{0.1cm}
\includegraphics[width=1in]{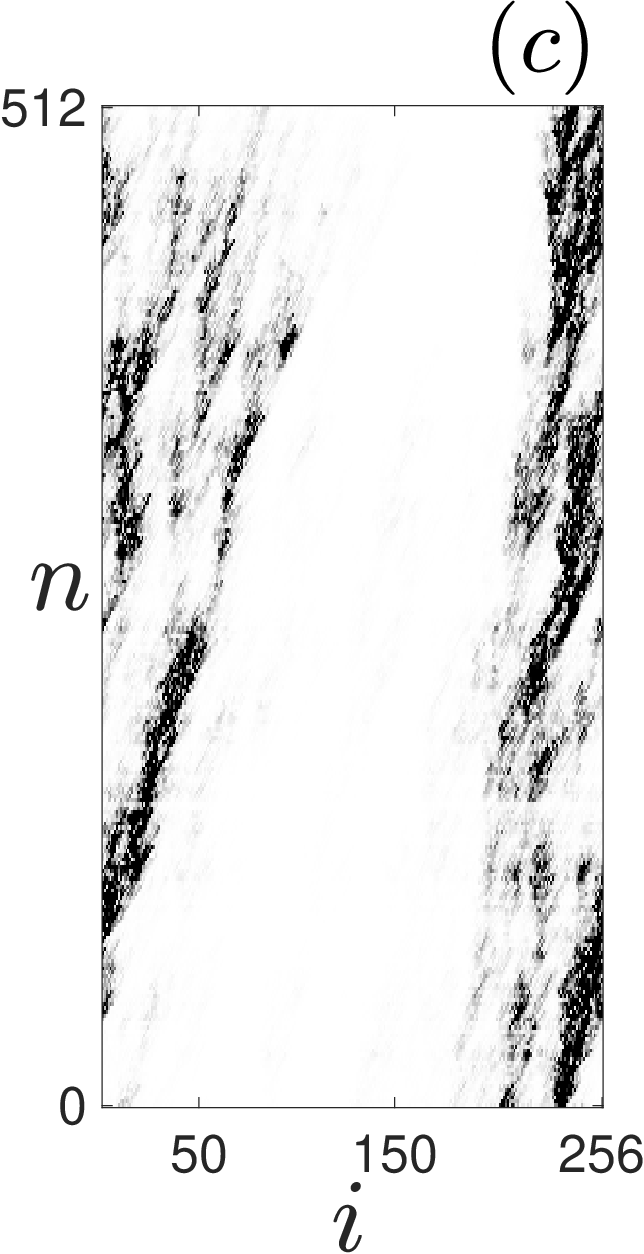}\\
\includegraphics[width=2.2in]{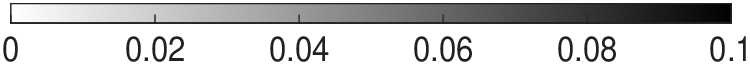}
\end{center}
\vspace{-9pt}
\caption{Spatiotemporal evolution of the magnitude of the leading CLV, $|\vec{v}_1^{(n)}|$, for:  ($a$)~$\alpha\!=\!0$, ($b$)~$\alpha\!=\!0.5$, and ($c$)~$\alpha\!=\!1$. Dark (light) contours are  large (small) magnitude. Regions of large magnitude tilt to the right with increasing $\alpha$ as a result of the drift velocity $v_d$ shown in Fig.~\ref{fig:vd}. Results are plotted using alternating time steps. Remaining parameters: $r\!=\!2.8$, $\epsilon\!=\!0.7$, $N \!=\!256$.}
\label{fig:clv1}
\end{figure}

The spatial structure of the CLVs provide insights into the growth or decay of small perturbations to the dynamics. Figure~\ref{fig:clv1} shows a spacetime plot of the magnitude of the leading CLV, $|\vec{v}_1^{(n)}|$ for three values of $\alpha$. Results are shown at alternating time steps to clearly visualize the larger structures that are present where dark (light) regions represent large (small) magnitude. 

In the absence of convective coupling, as shown in Fig.~\ref{fig:clv1}($a$), $|\vec{v}_1^{(n)}|$ is highly localized in space and travels vertically in the spacetime plot. For $\alpha \!>\! 0$, the CLV remains spatially localized while the spacetime structures drift with velocity $v_d$ as indicated by the tilting of regions of large magnitude towards the right as shown in Fig.~\ref{fig:clv1}($b$)-($c$). The CLVs can be characterized further by their spatial localization. Highly localized CLVs imply the presence of CLV components that have high amplitude in specific directions within the tangent space. High localization also indicates spatial regions on the lattice where the spatiotemporal dynamics are highly susceptible to local perturbations.

The spatial localization of the CLVs can be characterized using the inverse participation ratio~\cite{takeuchi:2011,mirlin:2000} defined as
$Y_2^{(k)} \!=\! \langle \sum_{i=1}^N | v_{k,i}^{(n)} |^4 \rangle$ 
where $v_{k,i}^{(n)}$ is the $i^{th}$ component of the $k^{th}$ CLV at time $n$. The angle brackets represent a time average over $4 \!\times\! 10^4$ time steps. The CLVs are normalized vectors with unit magnitude.

\begin{figure}
\begin{center}
\includegraphics[width=2.5in]{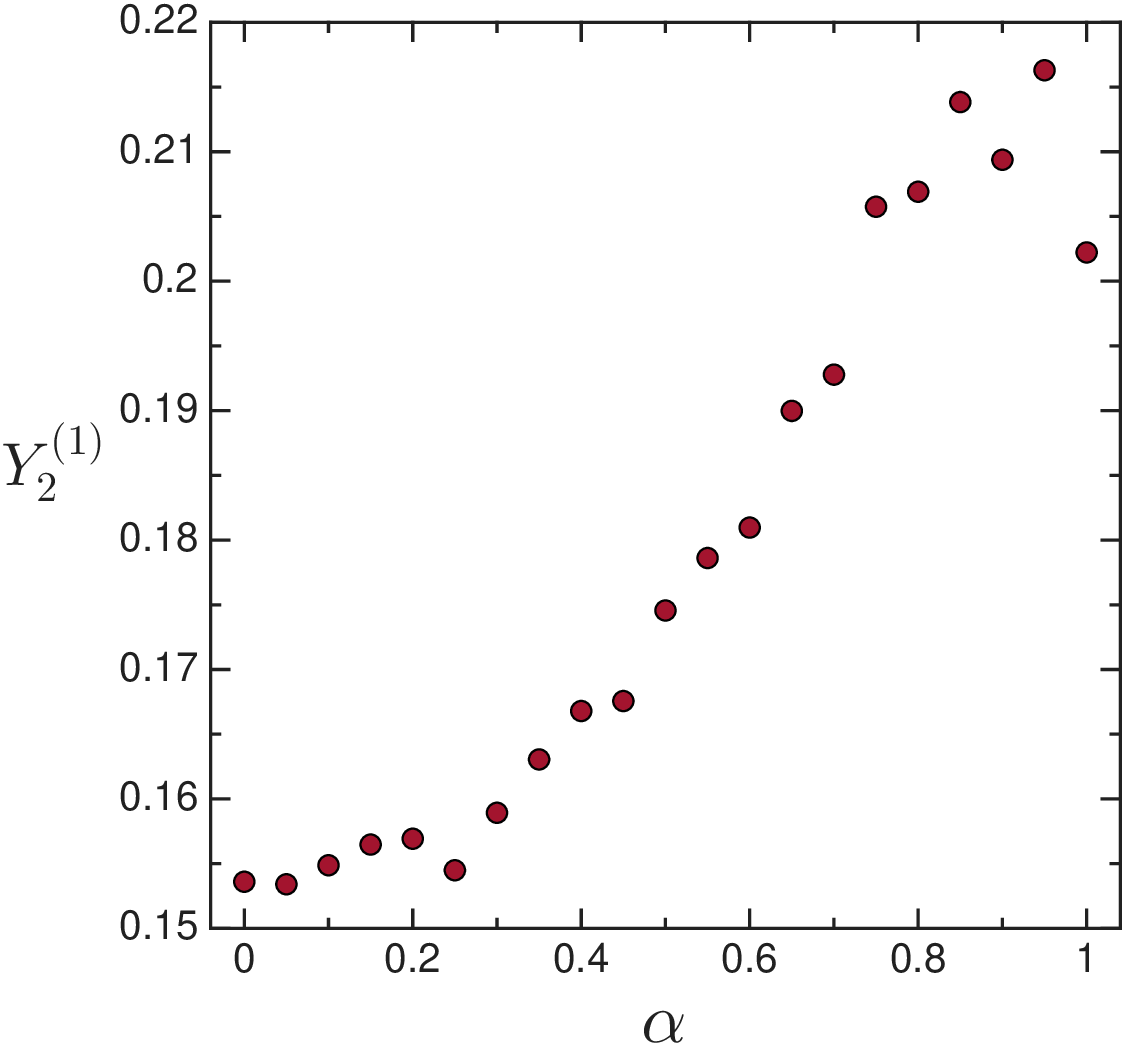}
\end{center}
\caption{The variation of the inverse participation ratio $Y_2^{(1)}$ of the leading CLV $\vec{v}_1$ with $\alpha$. Remaining parameters: $r\!=\!2.8$, $\epsilon\!=\!0.7$, $N \!=\!256$.}
\label{fig:loc1}
\end{figure}

Figure~\ref{fig:loc1} shows the variation of $Y_2^{(1)}$ with $\alpha$ over the range $0 \!\le\! \alpha \!\le\! 1$. The leading CLV is highly localized for all of the conditions we have explored. The general trend is increasing localization with increasing $\alpha$. These results indicate that large values of the convective coupling strength yield smaller spatial regions on the lattice that are highly susceptible to perturbations.

\begin{figure}
\begin{center}
\includegraphics[width=3.5in]{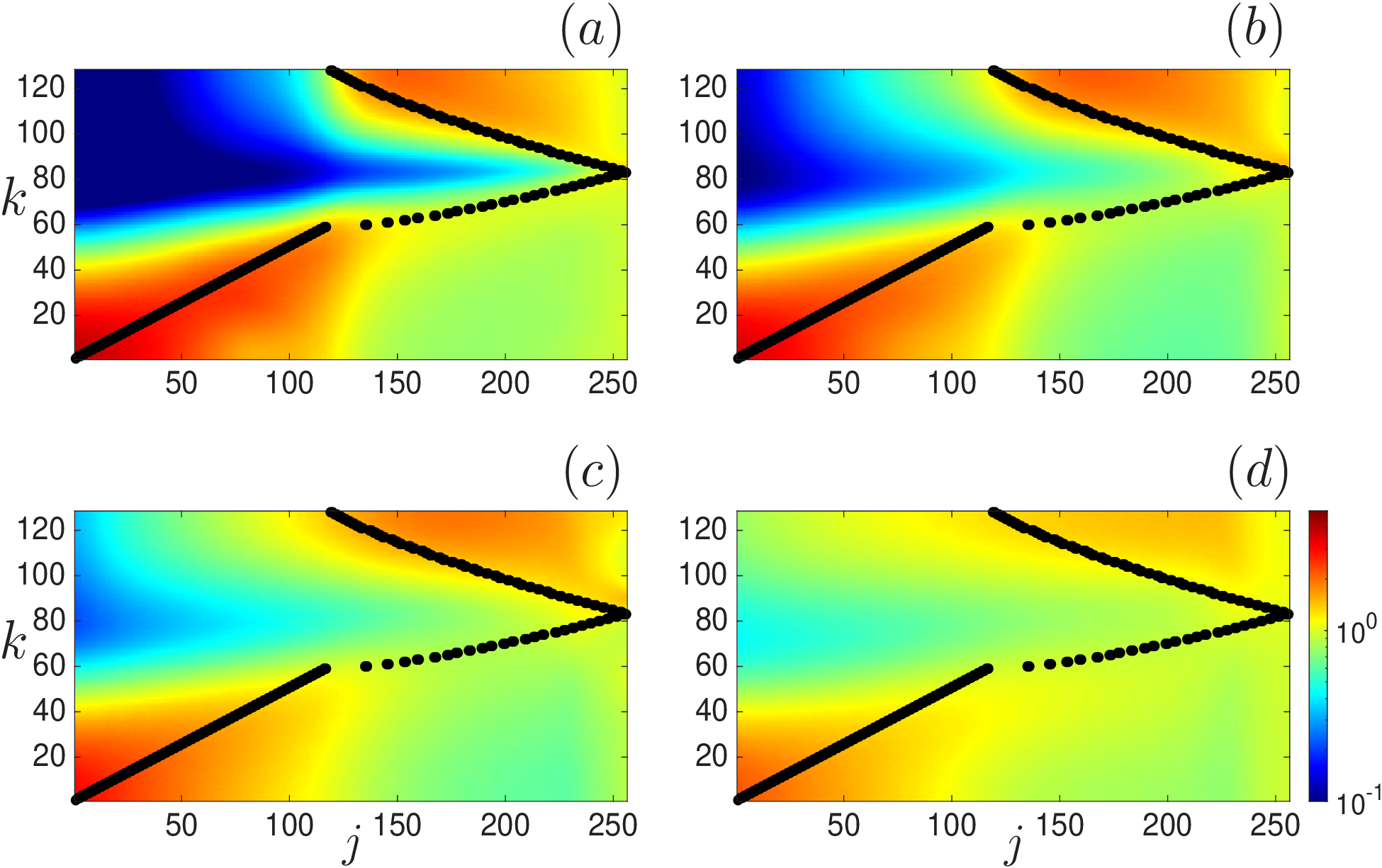}
\end{center}
\caption{Spatial power spectrum of the CLVs. Color contours of $\langle |\hat{v}_j|^2\rangle$ on a log scale, $j$ is the Lyapunov index, and $k$ is the integer wavenumber. Black circles are wavenumbers of the sorted eigenvectors of the diffusive coupling matrix $\mathbf{A_c}$. The region where the wavenumber of the eigenvectors grows linearly extends from $j\!=\!1$ to $j\!=\!117$. ($a$)~$\alpha\!=\!0.3$, ($b$)~$\alpha\!=\!0.5$, ($c$)~$\alpha\!=\!0.7$, ($d$)~$\alpha\!=\!1$. Remaining parameters: $r\!=\!2.8$, $\epsilon\!=\!0.7$, $N \!=\!256$.}
\label{fig:sps}
\end{figure}

The spatial structure of the CLVs is elucidated by computing the spatial power spectra~\cite{takeuchi:2011,raj:2024} as shown in Fig.~\ref{fig:sps} for four different values of $\alpha$. The horizontal axis is the Lyapunov index $j$ where $j \!=\! 1, 2, \!\ldots\!, N$. The vertical axis is the integer spatial wavenumber $k$ where $k \!=\! 1, 2, \!\ldots\!, N/2$. Color contours are the time-average of the magnitude of the spatial discrete Fourier transform of the CLV, $\langle\left|\hat{v}_j\right|^2\rangle$.   The hat notation represents a discrete Fourier transform in space and the angle brackets denote a time average. Black circles are the wavenumbers of the sorted eigenvectors of the diffusive coupling matrix $\mathbf{A_c}$. These eigenvectors are independent of $\alpha$ and their wavenumbers are included in every panel for comparison and to indicate the diffusive signature that is always present.

Figure~\ref{fig:sps}($a$) shows the spatial power spectrum of the CLVs for $\alpha = 0.3$. Significant power is localized in the lower left corner indicating that the leading CLVs (the CLVs with the largest Lyapunov exponents) have significant power at larger spatial structures. This coincides with the region where the wavenumber of the sorted eigenvectors increase linearly. More precisely, after the leading eigenvector, the eigenvectors in the linear regime occur in pairs with the wavenumber of the consecutive pairs. The linear region extends to a Lyapunov index of $j\!=\!117$ and is followed by a regime where the largest power occurs at very large wavenumber. In this regime, the linear trend of the eigenvectors is replaced with significant wavenumber mixing. 

A similar trend is seen in Fig.~\ref{fig:sps}($b$) for the spatial power spectrum when $\alpha \!=\! 0.5$. The eigenvalues of the diffusive coupling operator remain insightful. However, the regions of smallest power, indicated by dark blue, have diminished significantly. These trends continue for increasing $\alpha$ as shown in Fig.~\ref{fig:sps}($c$)-($d$). For  $\alpha\!=\!1$, the regions of very small power have disappeared while regions of significant power (shown in red) have remained. These results indicate that the convective coupling increases the power of the larger wavenumber structures of the CLVs. 

The variation of the magnitude exhibited on the spatial power spectra is quantified in Fig.~\ref{fig:sps} as a function of $\alpha$. The lower (upper) bound is the minimum (maximum) value of the power on the spatial power spectrum over all values of Lyapunov index $j$ and wavenumber $k$. For the case of diffusion alone ($\alpha=0$) the power ranges over seven orders of magnitude. The spread between the maximum and minimum decreases monotonically with its value becoming less than one order for $\alpha \!=\!1$.  Over the entire range of $\alpha$, the maximum value remains relatively constant while the minimum value increases significantly. This is evident in Fig.~\ref{fig:sps} by the significant reduction in low magnitude regions (blue) as $\alpha$ is increased.
\begin{figure}
\begin{center}
\includegraphics[width=3.25in]{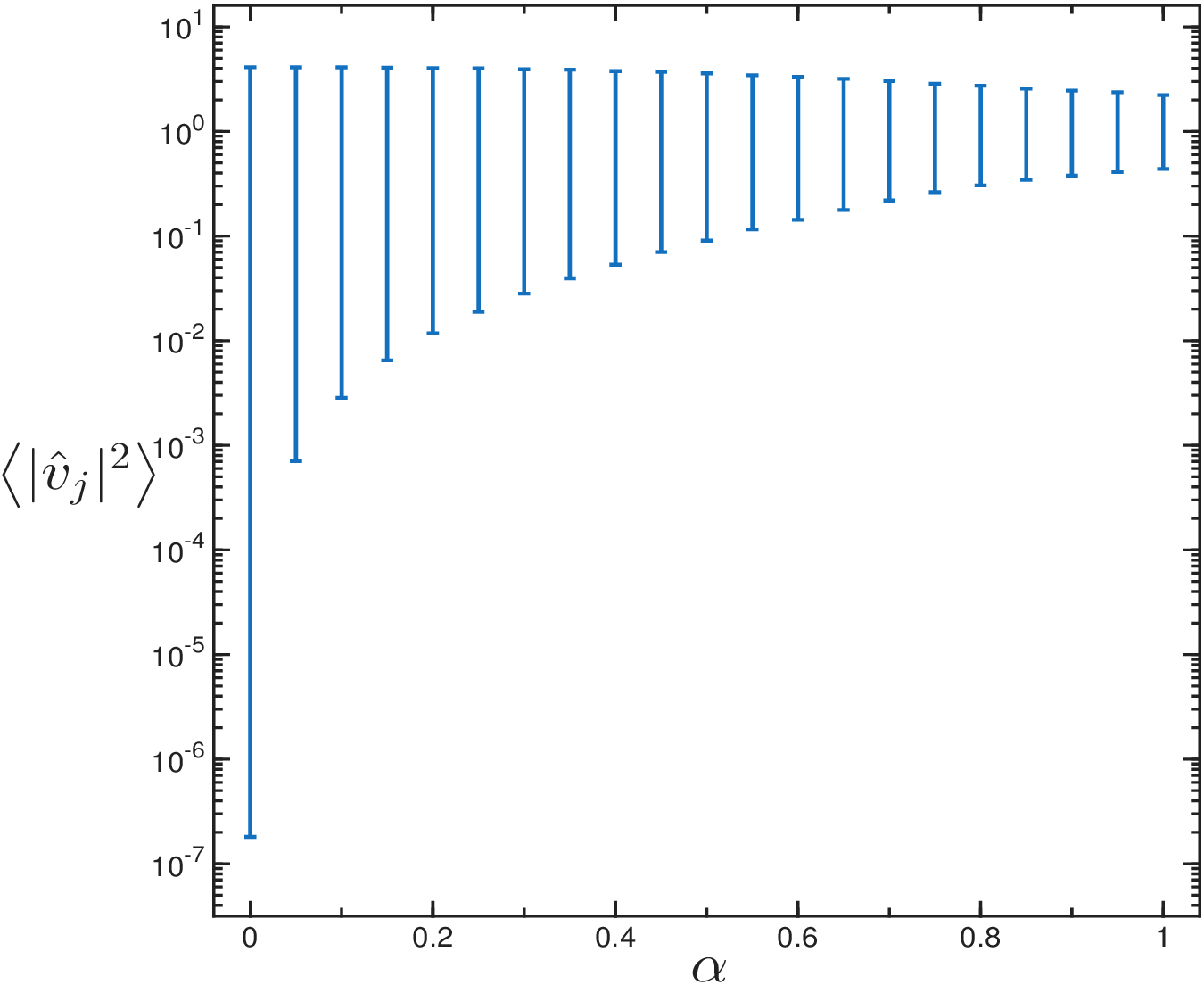}
\end{center}
\vspace{-9pt}
\caption{The range of values in the CLV spatial power spectra as a function of $\alpha$. The lower and upper bounds represent the minimum and maximum power found on the spatial power spectrum (for example, see Fig.~\ref{fig:sps}) for all wavenumber $k$ and Lyapunov index $j$ at that value of $\alpha$. Remaining parameters: $r\!=\!2.8$, $\epsilon\!=\!0.7$, $N \!=\!256$.}
\label{fig:sps-range}
\end{figure}

\subsection{Entangled CLVs}
\label{section:vdos}
According to Oseledets Multiplicative Ergodic Theorem~\cite{oseledec:1968}, the Lyapunov exponents $\lambda_k$ are guaranteed to be in descending order, $\lambda_1\!\ge\!\lambda_2\!\ge\!\ldots\!\ge\!\lambda_N$. This ordering, when applied to the finite-time covariant Lyapunov exponents, is often referred to as the domination of Oseledets splitting (DOS)~\cite{pikovsky:2016}. This strict ordering is not guaranteed to be true, for all time, for the fluctuating values of the finite time CLEs $\tilde{\lambda}_k$. Violations of the ordering occur when the CLVs become  nearly  tangent with each other and, as a result, their dynamics can become entangled. When CLVs are entangled, a perturbation along the direction of the nearly tangent CLVs can affect their dynamics. This must be contrasted with the hyperbolically isolated CLVs where a perturbation in the direction of one hyperbolically isolated CLV does not affect any of the other CLVs.

The fraction of time in which a violation of DOS occurs, $\nu_{k_1,k_2}^\tau$, is described by $\nu_{k_1,k_2}^\tau \!=\! \langle 1 \!-\! \mathcal{H} (\Delta \tilde{\lambda}_{k_1,k_2}^{\tau})\rangle_t$ where $\Delta \tilde{\lambda}_{k_1,k_2}^\tau \!=\! \tilde{\lambda}_{k_1} \!-\! \tilde{\lambda}_{k_2}$ is the difference between two finite time CLEs $\tilde{\lambda}_{k_1}$ and $\tilde{\lambda}_{k_2}$ averaged over the time interval $\tau$, and $k_1$ and $k_2$ are the Lyapunov indices of the two CLVs ~\cite{yang:2008,takeuchi:2011}. Averaging in time is indicated as $\langle \cdot \rangle_t$ and $\mathcal{H}$ is the Heaviside operator.  Many values of $\Delta \tilde{\lambda}_{k_1,k_2}^{\tau}$ are computed over long intervals of time, $t \!\gg\! \tau$. In our calculations, we use a total duration of $4 \!\times\! 10^4$ time steps with a finite time interval of $\tau\!=\!5$. We have explored using different values of $\tau$ and the results do not vary significantly with the findings we present here. DOS violations occur when $\Delta \tilde{\lambda}_{k_1,k_2}^{\tau} \!<\! 0$ for $k_1 \!>\! k_2$. For reference, $\nu_{k_1,k_2}^\tau \!=\! 1$ for CLVs that are always in violation and $\nu_{k_1,k_2}^\tau \!=\! 0$ for CLVs which are never in violation.
 
\begin{figure}
\begin{center}
\includegraphics[width=1.6in]{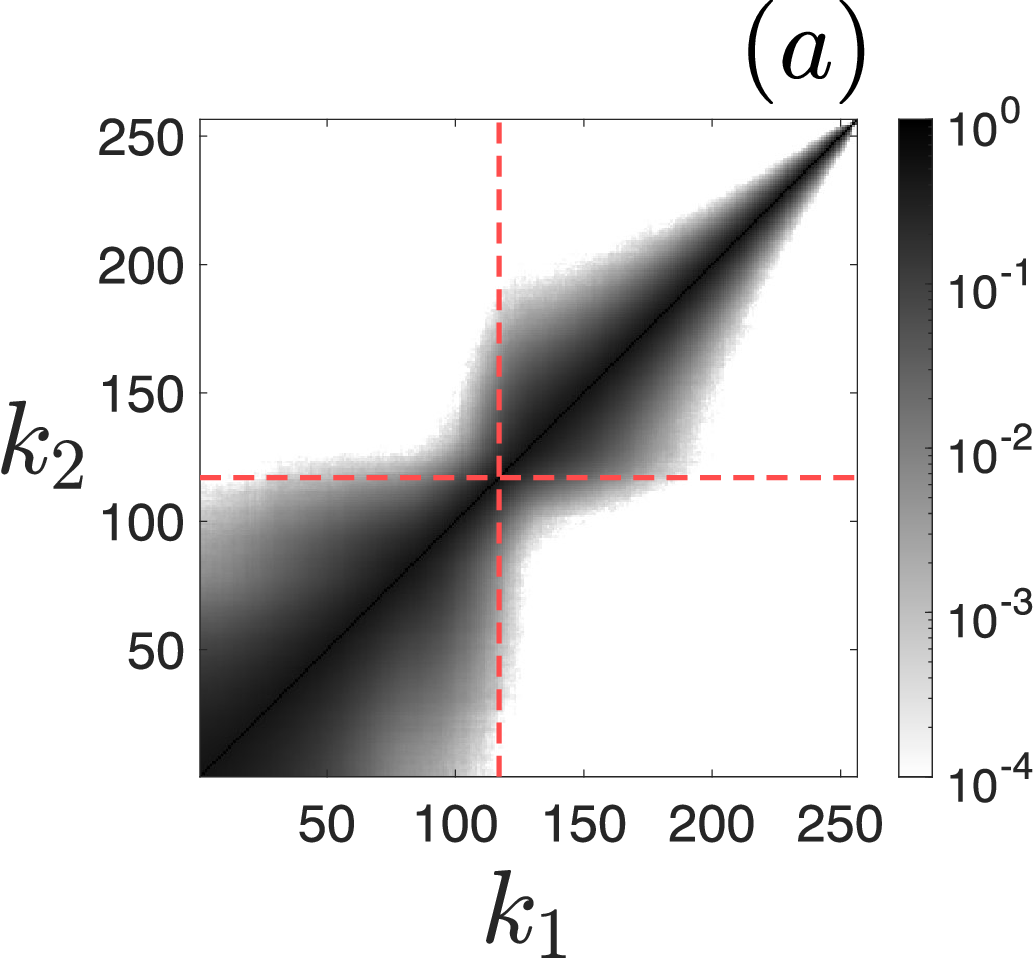} \hspace{0.1cm}
\includegraphics[width=1.6in]{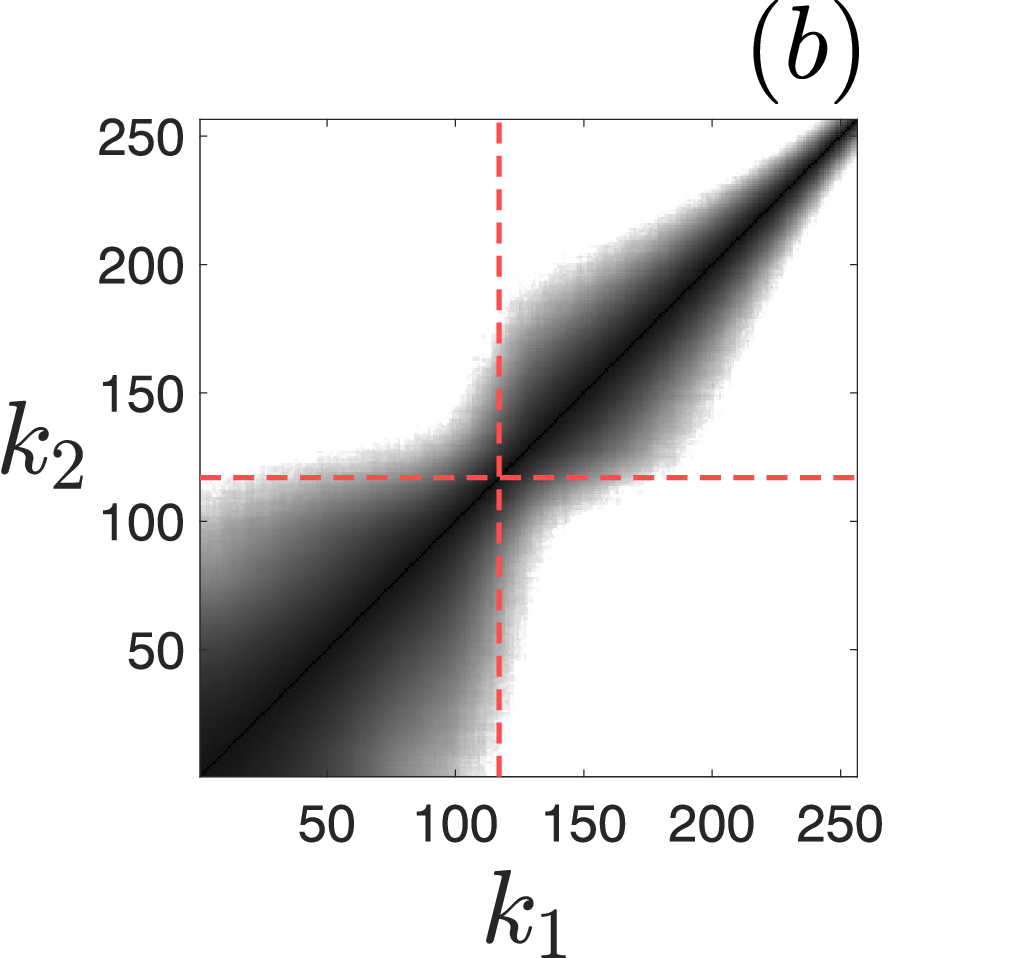} \\ \vspace{0.1cm}
\includegraphics[width=1.6in]{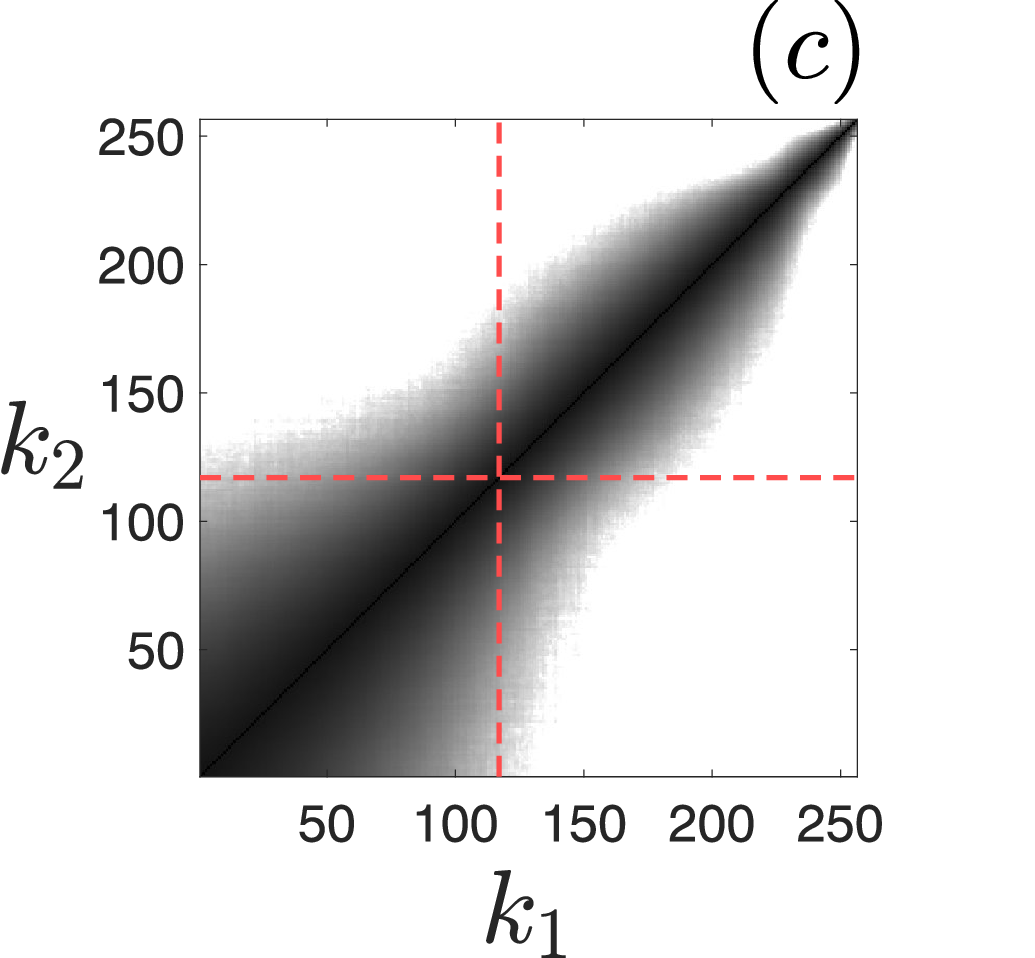} \hspace{0.1cm}
\includegraphics[width=1.6in]{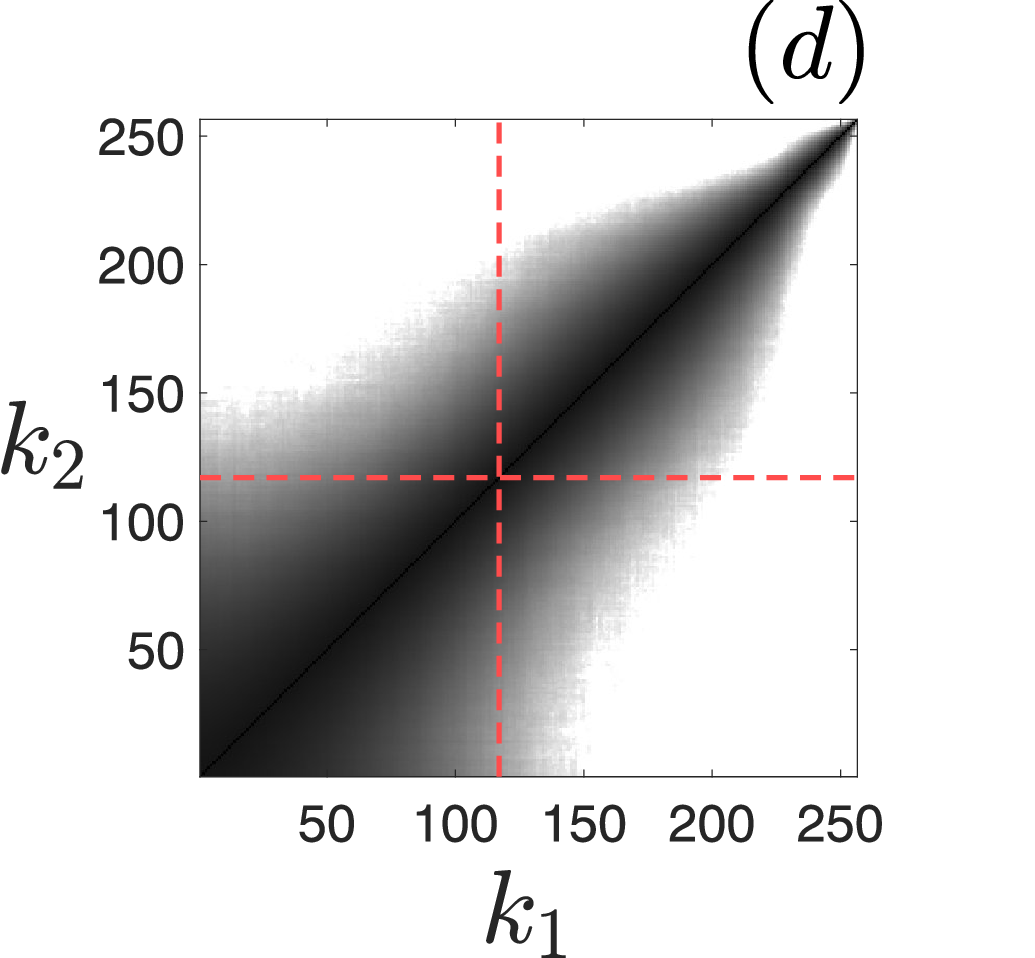}
\end{center}
\vspace{-9pt}
\caption{Violation of domination of Oseledets splitting between all pairs of CLVs where $k_1$ and $k_2$ are Lyapunov indices. Contours are shown for $\nu_{k_1,k_2}^\tau$ on a log scale using the color bar shown.  ($a$)~$\alpha=0$, ($b$)~$\alpha=0.3$, ($c$)~$\alpha=0.7$, ($d$)~$\alpha=1$. The red dashed lines represent $k_1\!=\!k_2\!=\!117$ corresponding to the linear regime for the eigenvectors of the diffusing coupling matrix (see Fig.~\ref{fig:sps}). Remaining parameters: $r\!=\!2.8$, $\epsilon\!=\!0.7$, $N \!=\!256$.}
\label{fig:vdos}
\end{figure}

DOS violations are shown for every pair of CLVs in Fig.~\ref{fig:vdos}.  Contours of $\nu_{k_1,k_2}^\tau$ are shown on a log scale where dark regions indicate a significant number of violations and light regions indicate the absence of violations. A dark line indicating pure violation occurs along the diagonal by $k_1\!=\!k_2$ as expected. These figures are symmetric about this diagonal by construction since every pixel represents a pairwise calculation where the order of the two Lyapunov indices not matter ($\nu_{k_1,k_2}^\tau\!=\!\nu_{k_2,k_1}^\tau)$.  Moving away from the diagonal it is clear that each CLV has significant violations with its neighboring CLVs. The amount of violation decreases as the difference between the indices increases.

The influence of the diffusive coupling matrix $\mathbf{A}_c$ is evident in the violations plotted in 
Fig.~\ref{fig:vdos}. The red dashed lines in Fig.~\ref{fig:vdos} correspond to the range of eigenvectors whose wavenumber grows linearly with index as shown in Fig.~\ref{fig:sps}.  The red dashed lines partition the violation of DOS in several interesting ways.

The case of diffusive coupling alone ($\epsilon \!>\!0$, $\alpha\!=\!0$) is shown in Fig.~\ref{fig:vdos}($a$). The violations within the lower region contained by red dashed lines ($k_1 \le k^*$ and $k_2 \le k^*$) are all significant. In addition, there are very few violations outside of this range in the region contained by $k_1 \!\le\! k^*$ and $k_2 \!>\! k^*$ (and by symmetry the region contained by $k_2 \!\le\! k^*$ and $k_1 \!>\! k^*$). Furthermore, the location on Fig.~\ref{fig:vdos} where $k_1\!=\!k_2\!=\!k^*$ represents a pinch point in the violations. 

Figure~\ref{fig:vdos}($b$) shows DOS violations for $\alpha\!=\!0.3$ where the violations are quite similar to the diffusion only case.  However, for larger values of $\alpha$ the amount of DOS violations around the pinch point increases significantly as shown in Fig.~\ref{fig:vdos}($c$)-($d$). The increased convective coupling is resulting in a significant increase in the number of DOS violations. This implies that CLV  entanglement increases with increasing $\alpha$.

It is interesting to note that all of the CLVs, for all of the conditions we explore here, are entangled physical modes~\cite{takeuchi:2011}. This is in agreement with what we found previously when studying diffusively coupled chaotic maps. However, our results presented here suggest that convective coupling increases the overall degree of entanglement of the CLVs. This is in agreement with the results found for Rayleigh-B\'enard convection~\cite{xu:2016} where the convective nonlinearity plays a significant role in the dynamics. For Rayleigh-B\'enard convection, 237 CLVs were computed using the full Boussinesq equations which were all found to be significantly entangled~\cite{xu:2016}.

\begin{figure}
\begin{center}
\includegraphics[width=3in]{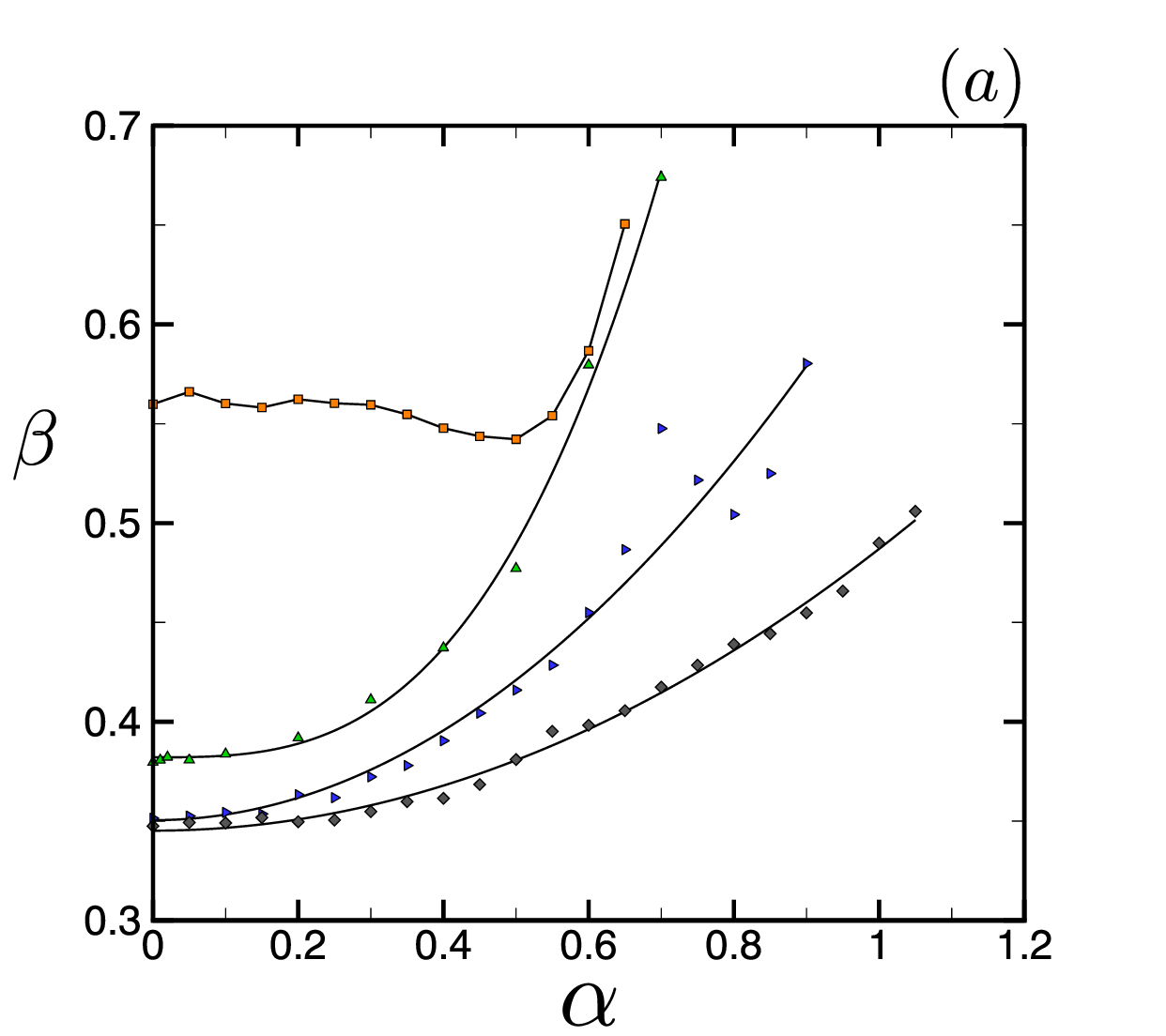}
\includegraphics[width=3in]{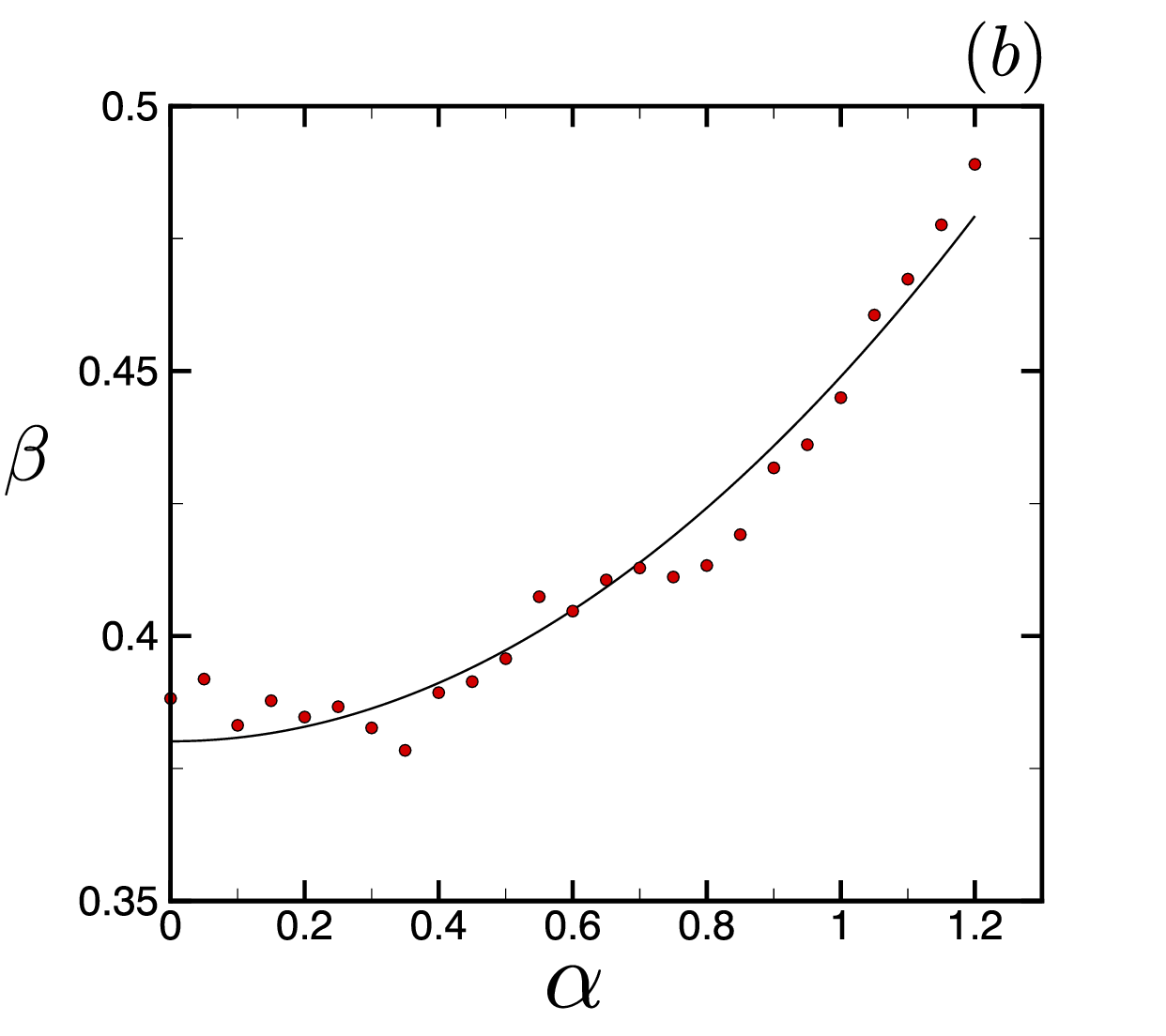}
\end{center}
\vspace{-9pt}
\caption{The variation of CLV entanglement, $\beta$, with $\alpha$ and $\epsilon$. ($a$) $\epsilon \!=\! 0.4$ (orange, squares); $\epsilon \!=\! 0.5$ (green, triangles) with fit $\beta \!=\! 0.38 \!+\! 0.86 \alpha^3$; $\epsilon \!=\! 0.6$ (blue, right triangles) with fit $\beta \!=\! 0.35 \!+\! 0.28 \alpha^2$, and $\epsilon \!=\! 0.7$ (gray, diamonds) with fit $\beta \!=\! 0.35 \!+\! 0.14 \alpha^2$. ($b$) $\epsilon \!=\! 0.8$ (red, circles) with fit $\beta \!=\! 0.38 \!+\! 0.069 \alpha^2$. Remaining parameters: $r\!=\!2.8$, $N \!=\!256$.}
\label{fig:entanglement}
\end{figure}

The amount of tangencies in the CLVs can be characterized by the entanglement $\beta$~\cite{barbish:2023} which quantifies the fraction of CLV pairs for which $\nu_{\tau}\!\geq\! 0.001$ (indicating violations of 0.1\% of the time or more). Figure~\ref{fig:entanglement} shows the entanglement as a function of $\alpha$ and $\epsilon$. 

Figure~\ref{fig:entanglement}($a$) shows results for $0.4 \!\le\! \epsilon \!\le\! 0.7$. The entanglement, for all values of $\alpha$, decreases with increasing $\epsilon$. For $\epsilon \!=\! 0.4$ the entanglement remains nearly constant for $\alpha \!\lesssim\! 0.6$ before increasing significantly for larger convective coupling strengths. For $0.5 \!\le\! \epsilon \!\le\! 0.7$, the entanglement increases smoothly with increasing $\alpha$, for these values of $\epsilon$. The solid lines are fits through the data where $\beta \!\sim\! \alpha^3$ for $\epsilon\!=\!0.5$ and $\beta \!\sim\! \alpha^2$ for $\epsilon\!=\!0.6$ and 0.7. We note that each curve is limited in the accessible values of $\alpha$ that yield numerically stable dynamics as shown by the upper edge of color in Fig.~\ref{fig:stability}.

The entanglement for $\epsilon \!=\! 0.8$ is shown separately in Fig.~\ref{fig:entanglement}($b$) since it overlaps with the curves shown in Fig.~\ref{fig:entanglement}($a$). The overall trend for $\epsilon \!=\! 0.8$ is an increasing entanglement with $\alpha$ as illustrated by the fit (solid line) with the form $\beta \sim \alpha^2$.

\section{Conclusion}
\label{section:conclusion}
We have examined the role of spatial coupling on the spatiotemporal chaos of large 1D lattices of convectively and diffusively coupled quadratic maps on a ring. The convective coupling has a significant impact upon the lattice dynamics and the spatiotemporal features of the CLVs.  For the conditions we explored, the convective coupling results in a drift velocity that carries spatial structures in the spacetime domain as well as regions of large CLV magnitude in the tangent space.

The leading Lyapunov exponent is not significantly sensitive to the spatial coupling over a large range of parameters while the shape of the spectrum of Lyapunov exponents changes significantly.  The overall trend with increasing convective coupling, is an increase in the Lyapunov exponents except for the large negative Lyapunov exponents which decrease. The overall result is a fractal dimension that increases significantly with increasing convective coupling strength. For example, diffusive coupling $\epsilon=0.7$ in the absence of convection, $\alpha\!=\!0$, yields a fractal dimension of $D_\lambda \!=\! 138$. For the same conditions with $\alpha\!=\!1$ the fractal dimension increases to $D_\lambda \!=\! 203$ representing a 50\% increase. This suggests the importance of convective coupling on the chaotic dynamics of fluid transport problems where the value of the fractal dimension is expected to be large.

The convective coupling increases the entanglement of the CLVs. In all of our results, all of the CLVs are entangled with their neighboring CLVs. Furthermore, the entanglement increases rapidly as the convective coupling strength is increased. These findings are in alignment with what has been found for the chaotic dynamics of a fluid layer undergoing Rayleigh-B\'enard convection where convective coupling is significant. These results suggest that dynamics with convective coupling may have elevated values of the fractal dimension with highly entangled CLVs.

~\vspace{0.2cm}

\noindent Acknowledgments: We acknowledge support from NSF fund number CMMI-2138055. Portions of the computations were performed using the Advanced Research Computing (ARC) center at Virginia Tech.

\bibliographystyle{unsrt}

\end{document}